\pdfoutput=1
\documentclass[conference]{IEEEtran}
\IEEEoverridecommandlockouts

\usepackage{cite}
\usepackage{amsmath,amssymb,amsfonts}
\usepackage{graphicx}
\usepackage{textcomp}
\usepackage{xcolor}
\usepackage{booktabs}
\usepackage[hyphens]{url}
\usepackage{hyperref}

\newcommand{\AnchorBelowPTenPctTwentyThree}{27}
\newcommand{\AnchorBumpedPctTwentyThree}{44}
\newcommand{\AnchorMedRatioMax}{1.21}
\newcommand{\AnchorMedRatioMin}{0.96}

\newcommand{\AnchorVbMed}{639}
\newcommand{\AnchorsAll}{387{,}502}

\newcommand{\BillBtcTwentySix}{1.5}
\newcommand{\BillBtcTwentyThree}{16.8}

\newcommand{\BillFeePctMax}{0.18}
\newcommand{\BillFeePctMin}{0.03}

\newcommand{\BillVbPctMax}{0.12}

\newcommand{\BreachLookAlikesWeek}{90}
\newcommand{\BreachOpenToSpliceMed}{5}

\newcommand{\BreachSpliceToCloseMed}{3}
\newcommand{\BreachWaveBtc}{3.18}
\newcommand{\BreachWaveChannels}{305}
\newcommand{\BreachWavePenalised}{306}
\newcommand{\BreakEven}{6.38}
\newcommand{\BreakEvenRateAll}{16.7}
\newcommand{\BreakEvenRateEarlyMax}{54.1}
\newcommand{\BreakEvenRateEarlyMin}{6.1}
\newcommand{\BreakEvenRateTwentyFive}{2.7}
\newcommand{\BreakEvenRateTwentySix}{1.2}
\newcommand{\BtcDaysLockedAll}{68{,}036}
\newcommand{\BumpPctHtlcTwentyThree}{78}
\newcommand{\BumpPctNoHtlcTwentyThree}{32}
\newcommand{\BumpShareTwentyThree}{34}
\newcommand{\CascadeAllTimeoutPct}{96}

\newcommand{\CascadeMaxCloses}{18}
\newcommand{\CascadePayments}{6{,}951}
\newcommand{\CascadePct}{25}

\newcommand{\CascadeSpanMed}{140}
\newcommand{\CfAnchorBtc}{1.28}
\newcommand{\CfAnchorVb}{173}

\newcommand{\CfCloses}{223{,}587}
\newcommand{\CfNotBumpedPct}{88}
\newcommand{\CfPtwoaVb}{152}
\newcommand{\CfSavingPct}{12}
\newcommand{\CltvDeltaFrom}{May~2023}
\newcommand{\CohortOneYearNineteen}{35}

\newcommand{\CohortOneYearTwentyFive}{17}

\newcommand{\CohortOneYearTwentyOne}{24}
\newcommand{\CohortOneYearTwentyThree}{31}

\newcommand{\CostMedTwenty}{14{,}318}

\newcommand{\CostMedTwentySix}{925}
\newcommand{\CostMedTwentyThree}{7{,}346}

\newcommand{\CostRatioMax}{6}
\newcommand{\CostRatioMin}{2}

\newcommand{\CostTotalBtcAll}{31.1}

\newcommand{\CostTotalBtcTwentyThree}{11.97}

\newcommand{\CutoffDate}{31~October~2026}
\newcommand{\CutoffHeight}{974{,}041}
\newcommand{\DelayOneDayPct}{49}
\newcommand{\DelayTwoWeeksPct}{10}

\newcommand{\ElastUnlisted}{-0.22}

\newcommand{\ForceShareNineteen}{50}

\newcommand{\ForceShareTwentySix}{20}
\newcommand{\ForceShareTwentyThree}{44}

\newcommand{\HtlcCommitBeforeExpiryPct}{56}
\newcommand{\HtlcLateEarlyMax}{8.0}
\newcommand{\HtlcLateEarlyMin}{6.5}

\newcommand{\HtlcLateMidAnchor}{4.7}
\newcommand{\HtlcLateMidLegacy}{13.2}

\newcommand{\HtlcLateSpikeAnchor}{20.6}
\newcommand{\HtlcLateSpikeLegacy}{30.5}

\newcommand{\HtlcLateTwentyFour}{16.4}
\newcommand{\HtlcLateTwentyFourAnchor}{12.8}
\newcommand{\HtlcLateTwentyFourLegacy}{6.3}

\newcommand{\HtlcLateTwentyThree}{18.2}

\newcommand{\HtlcMarginMid}{80}
\newcommand{\HtlcMarginTight}{40}

\newcommand{\HtlcShareTwentyThree}{14}
\newcommand{\HtlcSuccessLateMaxOther}{2.4}
\newcommand{\HtlcSuccessLateMinOther}{1.0}
\newcommand{\HtlcSuccessLateNineteen}{8.6}
\newcommand{\HtlcSuccessLateTwentyFour}{16.0}
\newcommand{\HtlcSuccessLateTwentyThree}{8.5}
\newcommand{\HtlcSuccessMarginMax}{19}
\newcommand{\HtlcSuccessMarginMin}{8}

\newcommand{\HtlcTimeoutMedMax}{3}
\newcommand{\HtlcTimeoutMedMin}{1}
\newcommand{\HtlcTimeoutTxs}{70{,}808}

\newcommand{\LcCapPct}{0.15}
\newcommand{\LcCloseMed}{1{,}411}
\newcommand{\LcCoopMed}{1{,}970}
\newcommand{\LcForceCoopRatio}{3.4}
\newcommand{\LcForceCoopRatioMax}{3.9}
\newcommand{\LcForceCoopRatioMin}{1.3}
\newcommand{\LcForceMed}{6{,}645}

\newcommand{\LcLifeDaysMed}{75}
\newcommand{\LcMed}{3{,}435}
\newcommand{\LcMedEarlyRegime}{10{,}510}
\newcommand{\LcMedHighRegime}{6{,}483}
\newcommand{\LcMedLowRegime}{986}

\newcommand{\LcOpenBtc}{28.2}
\newcommand{\LcOpenMed}{723}
\newcommand{\LcOpensPct}{40}
\newcommand{\LcOpensPctLowRegime}{63}

\newcommand{\LcRobustBatchPct}{1}
\newcommand{\LcRobustMaxPct}{19}
\newcommand{\LcRobustMinPct}{18}

\newcommand{\LcTotalBtc}{70.9}
\newcommand{\LegacyClusterAgeDays}{4}
\newcommand{\LegacyMedRatioEarlyMax}{1.3}
\newcommand{\LegacyMedRatioEarlyMin}{1.1}

\newcommand{\LegacyMedRatioTwentySix}{2.0}

\newcommand{\LegacyOldAgeYears}{3.9}

\newcommand{\LegacyOverBtcAll}{10.9}

\newcommand{\LegacyOverTwoPctTwentyOne}{32}
\newcommand{\LegacyOverTwoPctTwentySix}{52}
\newcommand{\LegacyPctExClusterTwentySix}{8}

\newcommand{\LegacyTopFivePct}{74}
\newcommand{\LegacyVbMed}{292}
\newcommand{\MassEventsMax}{239}
\newcommand{\MassEventsMin}{64}
\newcommand{\MassLargest}{856}
\newcommand{\MassMin}{20}
\newcommand{\MassShareMax}{32}
\newcommand{\MassShareMin}{19}

\newcommand{\MinersShareLateMax}{87}
\newcommand{\MinersShareLateMin}{83}

\newcommand{\MinersShareTwentySix}{87}

\newcommand{\MinersShareTwentyTwo}{24}

\newcommand{\MutualBtcTwentyFive}{19{,}796}
\newcommand{\MutualBtcTwentyFour}{8{,}885}

\newcommand{\NClosesAll}{973{,}581}
\newcommand{\NClosesTitle}{974{,}000}
\newcommand{\NForceAll}{523{,}815}
\newcommand{\NFundingOutpoints}{736{,}110}
\newcommand{\NOutputs}{1{,}346{,}588}
\newcommand{\NOutputsSpent}{1{,}212{,}596}
\newcommand{\NPublicCloses}{699{,}049}
\newcommand{\NPublicForce}{249{,}283}
\newcommand{\NPublicMutual}{449{,}707}
\newcommand{\NSpikes}{9}
\newcommand{\NSpikesDown}{5}
\newcommand{\NSpikesUp}{4}
\newcommand{\NUnlistedForce}{274{,}532}

\newcommand{\NotRaisedPctTwentySix}{70}
\newcommand{\OpenAtCutoff}{40{,}353}
\newcommand{\OpenJulTwentySix}{50{,}219}
\newcommand{\OpenPeak}{90{,}340}
\newcommand{\OrganicRateNineteen}{50}

\newcommand{\OrganicRateTwentySix}{26}
\newcommand{\OrganicRateTwentyThree}{41}

\newcommand{\OverOnePctCapTwentyOne}{41}
\newcommand{\OverOnePctCapTwentySix}{8}

\newcommand{\OwnerPctTwentySix}{5}

\newcommand{\OwnerPctTwentyTwo}{81}

\newcommand{\PairClosesLdkLdk}{13{,}415}

\newcommand{\PairClosesLndLnd}{169{,}435}

\newcommand{\PairForcePctClnCln}{40}
\newcommand{\PairForcePctClnLnd}{35}
\newcommand{\PairForcePctEclairLnd}{29}

\newcommand{\PairForcePctLdkLdkExact}{1.5}
\newcommand{\PairForcePctLdkLnd}{37}
\newcommand{\PairForcePctLndLnd}{28}
\newcommand{\PairForcePctLndUnknown}{44}
\newcommand{\PairLdkLdkTopNodePct}{99.5}
\newcommand{\PenaltyBtcAll}{11.4}
\newcommand{\PenaltyMonthlyMaxBefore}{61}
\newcommand{\PenaltyMonthlyMaxBeforeMonth}{September~2018}
\newcommand{\PenaltyMonthlyMed}{2}
\newcommand{\PenaltyTxsAll}{950}
\newcommand{\PenaltyTxsTwentySix}{321}

\newcommand{\PenaltyWaveFirst}{28~September~2026}
\newcommand{\PenaltyWaveLast}{5~October~2026}
\newcommand{\PenaltyWaveMaxBlocks}{120}

\newcommand{\PlBlockDraws}{5{,}000}
\newcommand{\PlBlockWeeks}{8}
\newcommand{\PlDetectableWavePct}{15}
\newcommand{\PlEvAllLater}{1.10}
\newcommand{\PlEvAllLaterHi}{1.14}
\newcommand{\PlEvAllLaterLo}{0.90}
\newcommand{\PlEvAllLaterMed}{1.01}

\newcommand{\PlEvAllLaterPct}{88}
\newcommand{\PlEvAllNow}{0.97}
\newcommand{\PlEvAllNowHi}{1.17}
\newcommand{\PlEvAllNowLo}{0.90}
\newcommand{\PlEvAllNowMed}{1.02}
\newcommand{\PlEvAllNowPct}{23}

\newcommand{\PlEvAnchorNow}{0.94}
\newcommand{\PlEvAnchorNowHi}{1.26}
\newcommand{\PlEvAnchorNowLo}{0.86}
\newcommand{\PlEvAnchorNowMed}{1.04}
\newcommand{\PlEvAnchorNowPct}{17}

\newcommand{\PlEvLegacyNow}{0.98}
\newcommand{\PlEvLegacyNowHi}{1.19}
\newcommand{\PlEvLegacyNowLo}{0.89}
\newcommand{\PlEvLegacyNowMed}{1.00}
\newcommand{\PlEvLegacyNowPct}{42}
\newcommand{\PlFwPBlock}{0.359}
\newcommand{\PlFwPCirc}{0.420}
\newcommand{\PlLagFourB}{0.00}

\newcommand{\PlLagFourBlockPct}{51}
\newcommand{\PlLagFourCircHi}{0.07}
\newcommand{\PlLagFourCircLo}{-0.07}

\newcommand{\PlLagFourCircPct}{46}
\newcommand{\PlLagOneB}{0.01}

\newcommand{\PlLagOneBlockPct}{63}
\newcommand{\PlLagOneCircHi}{0.08}
\newcommand{\PlLagOneCircLo}{-0.08}

\newcommand{\PlLagOneCircPct}{61}
\newcommand{\PlLagThreeB}{0.01}

\newcommand{\PlLagThreeBlockPct}{69}
\newcommand{\PlLagThreeCircHi}{0.07}
\newcommand{\PlLagThreeCircLo}{-0.08}

\newcommand{\PlLagThreeCircPct}{65}
\newcommand{\PlLagTwoB}{0.02}

\newcommand{\PlLagTwoBlockPct}{76}
\newcommand{\PlLagTwoCircHi}{0.07}
\newcommand{\PlLagTwoCircLo}{-0.08}

\newcommand{\PlLagTwoCircPct}{67}
\newcommand{\PlLeadOneB}{0.02}

\newcommand{\PlLeadOneBlockPct}{72}
\newcommand{\PlLeadOneCircHi}{0.07}
\newcommand{\PlLeadOneCircLo}{-0.08}

\newcommand{\PlLeadOneCircPct}{69}
\newcommand{\PlMaxLagB}{0.02}

\newcommand{\PlMaxLagBlockPct}{64}
\newcommand{\PlMaxLagCircHi}{0.08}
\newcommand{\PlMaxLagCircLo}{-0.06}

\newcommand{\PlMaxLagCircPct}{58}
\newcommand{\PlMinShift}{13}
\newcommand{\PlNwRejectBlockMax}{14}
\newcommand{\PlNwRejectBlockMin}{12}
\newcommand{\PlNwRejectCircMax}{25}
\newcommand{\PlNwRejectCircMin}{21}
\newcommand{\PlSameWeekB}{0.00}

\newcommand{\PlSameWeekBlockPct}{55}
\newcommand{\PlSameWeekCircHi}{0.07}
\newcommand{\PlSameWeekCircLo}{-0.08}

\newcommand{\PlSameWeekCircPct}{55}
\newcommand{\PlSeed}{42}
\newcommand{\PlShifts}{382}

\newcommand{\PlWeekDraws}{10{,}000}

\newcommand{\PubForceTwentySix}{15{,}440}

\newcommand{\PubForceTwentyTwo}{57{,}896}

\newcommand{\PubLegacyPctTwentyOne}{87}
\newcommand{\PubLegacyPctTwentySix}{26}

\newcommand{\RateAugTwentySix}{51}
\newcommand{\RateOrganicAugTwentySix}{31}
\newcommand{\RateOrganicFebNineteen}{106}
\newcommand{\RateOrganicMarNineteen}{103}
\newcommand{\RateOrganicMaxLater}{64}

\newcommand{\ScanStartHeight}{503{,}816}
\newcommand{\SepEventBlocks}{18}
\newcommand{\SepEventBtc}{36.7}
\newcommand{\SepEventCloses}{8{,}990}
\newcommand{\SepEventHours}{2}
\newcommand{\SepEventLegacy}{6{,}910}
\newcommand{\SpikeAprTwentyFourPct}{50}
\newcommand{\SpikeMarNineteenPct}{26}
\newcommand{\SpikeMayTwentyThreeFee}{84}
\newcommand{\SpikeMayTwentyThreeFeeBefore}{13}
\newcommand{\SpikeMayTwentyThreePct}{29}
\newcommand{\SpikeMinFee}{10}

\newcommand{\SpikeRule}{2}
\newcommand{\SplitResBtc}{0.48}
\newcommand{\SplitResForwarderBtc}{0.44}
\newcommand{\SplitResForwarderPayments}{38}
\newcommand{\SplitResFromYear}{2021}
\newcommand{\SplitResGapMed}{58}
\newcommand{\SplitResPayments}{40}
\newcommand{\SplitResPreimageFirstPct}{100}
\newcommand{\SplitResStagedBtc}{0.0016}
\newcommand{\SplitResStagedPairs}{7}
\newcommand{\SplitResStagedPreimageAfterMed}{4}
\newcommand{\SplitResStagedToSelfDelay}{6}
\newcommand{\SweepBurnPct}{10}

\newcommand{\SweepCheaperPctTwentyThree}{85}
\newcommand{\SweepLargestFirst}{27~July~2024}
\newcommand{\SweepLargestNetBtc}{0.070}
\newcommand{\SweepLargestPct}{41}

\newcommand{\SweepLargestPctTwentySix}{58}

\newcommand{\SweepMedFrTwentyTwo}{1.0}
\newcommand{\SweepMinersBtc}{0.75}
\newcommand{\SweepOperators}{332}
\newcommand{\SweepOperatorsOverOne}{15}
\newcommand{\SweepPFiftyAtSixteenTwentyThree}{13}
\newcommand{\SweepPFiftyAtSweepTwentyThree}{5}
\newcommand{\SweepSweepersBtc}{0.20}

\newcommand{\SweepValueBtc}{0.95}
\newcommand{\TaprootAnchors}{23{,}864}
\newcommand{\TaprootAnchorsSpent}{1{,}706}

\newcommand{\TaprootAnchorsSpentTwentySix}{71}
\newcommand{\TaprootAnchorsTwentySix}{20{,}177}
\newcommand{\TaprootAnchorsUnspent}{22{,}158}

\newcommand{\TaprootCosweepToRemotePctEarly}{96}
\newcommand{\TaprootCosweepToRemotePctLate}{0.2}
\newcommand{\TaprootKeyExposedDaysMed}{101}
\newcommand{\TaprootKeyExposedUnspent}{14{,}828}

\newcommand{\TaprootKeyPublicBtc}{0.049}

\newcommand{\TaprootOwnerSweepFrEarly}{2.0}
\newcommand{\TaprootOwnerSweepFrLate}{0.9}

\newcommand{\TaprootScriptPathChainwide}{0}

\newcommand{\ThirdMedDelayDaysTwentyThree}{12}
\newcommand{\ThirdMedDelayTwentyFive}{16}

\newcommand{\ThirdMedDelayTwentyThree}{1{,}708}

\newcommand{\ToLocalMedDays}{1.7}
\newcommand{\ToLocalPNinetyMax}{22}

\newcommand{\UnlOverPubTwentySix}{3.4}
\newcommand{\UnlTaprootPctTwentySix}{23}

\newcommand{\ValAgreePct}{100.00}
\newcommand{\ValCompared}{480{,}440}
\newcommand{\ValCoveragePct}{69}
\newcommand{\ValCoverageYearMaxPct}{94}
\newcommand{\ValCoverageYearMinPct}{69}

\newcommand{\ValLakeCompared}{698{,}476}

\newcommand{\VbRatioMax}{3.1}
\newcommand{\VbRatioMin}{3.0}
\newcommand{\ZfBeforeMerge}{6}

\newcommand{\ZfCloses}{31}
\newcommand{\ZfFirst}{29~October~2025}
\newcommand{\ZfFirstPublic}{28~September~2026}
\newcommand{\ZfMedCap}{49{,}155}
\newcommand{\ZfPkgFr}{1.6}
\newcommand{\ZfPublic}{13}
\newcommand{\ZfPublicNodes}{10}
\newcommand{\ZfUnlisted}{18}

\newcommand{\PaperTitle}{What Does It Cost to Force-Close a Lightning Channel?}
\newcommand{\PaperSubtitle}{Evidence from \NClosesTitle{} Channel Closes, 2018--2026}

\hypersetup{pdftitle={\PaperTitle{} \PaperSubtitle}, pdfauthor={Saurabh Kumar}}

\begin{document}

\title{\PaperTitle\\
{\LARGE \PaperSubtitle}}

\author{
\IEEEauthorblockN{Saurabh Kumar}
\IEEEauthorblockA{Bitcoin Data Labs\\
sorukumar@bitcoindatalabs.org}
}

\maketitle
\thispagestyle{plain}
\pagestyle{plain}

\begin{abstract}
Bitcoin's Lightning payment network returns the funds of a channel that cannot close cooperatively through
pre-signed transactions whose fees and delays fall on its operators. That fee design is being replaced: Bitcoin
Core 31.0 removed the mempool rule anchor channels relied on, and the Lightning specification adopted zero-fee
commitments in May~2026. What the old design costs has never been measured across fee regimes.
We trace every public channel close and every force close on-chain from January~2018 to October~2026,
\NClosesAll{} closes, and follow each output to its spend. A separate chain index confirms all \ValLakeCompared{}
public closes we could compare.
Fee spikes did not bring a wave of force closes. Force closes in a spike week and the next were \PlEvAllNow{}
times the four weeks before, against \PlEvAllNowMed{} after ordinary weeks of the same years.
A force close costs \CostRatioMin--\CostRatioMax{} times a cooperative close in fees and locks the closer's balance
for days, which outweighs the fees only in low-fee years. When fees are high, pending payments are exposed: in 2023,
\HtlcLateTwentyThree\% of HTLC timeouts confirmed more than \HtlcMarginTight{} blocks after expiry. Yet since
\SplitResFromYear, forwarding nodes lost only \SplitResForwarderPayments{} payments, each with the preimage already
on-chain. Anchor outputs cut late timeouts in fee spikes (\HtlcLateSpikeAnchor\% against
\HtlcLateSpikeLegacy\%), yet owners rarely bump commitments with them, and miners take most of what third parties
sweep. Zero-fee commitments would save only \CfSavingPct\% of the anchor mechanism's block space; they should be
judged by how promptly payments resolve when fees spike.

\end{abstract}


\section{Introduction}
\label{sec:intro}

The Lightning Network is a payment network built on Bitcoin. Its nodes hold their operators' funds in two-party
payment channels, each backed by one output on the Bitcoin chain. When both parties agree, a channel closes
cooperatively with one ordinary transaction. When they cannot, because a peer is offline, a payment is stuck or the two
implementations disagree, either party can force-close the channel: it broadcasts the latest commitment transaction,
signed in advance by both, and then claims its share through further pre-signed and time-locked
transactions~\cite{poon2016bitcoin,bolt3,bolt5}. The fees and delays of those transactions fall on the operators, and
a transaction that confirms too late can lose a pending payment or let a cheating peer
win~\cite{harris2020flood,riard2020timedilation,riard2023replacement}. The fee design of these transactions is
changing now. Bitcoin Core 31.0 (April~2026) removed the CPFP carve-out, the mempool rule that anchor channels were
designed around~\cite{core31}, and in May~2026 the Lightning specification adopted zero-fee commitments, which rely on
newer relay rules instead~\cite{bolts1228zerofee,bip431,core28}. This paper measures the old design, across the fee
regimes it lived through, before the migration, as a baseline for judging the new one.

Force closes have been counted~\cite{bitmex2020closures,zabka2022empirical,grotschla2025lifecycle} and
predicted~\cite{antonelli2026closures}, but what they cost has not been measured across fee regimes. The questions
that shaped the design are argued in developer forums from a few nodes' logs or from none: whether owners use their
anchor outputs to bump fees, whether third parties sweep them at a profit, and whether fee spikes force channels
closed~\cite{daftuar2024v3,lnd8997}. Our previous paper measured the cost of force closes over a few months of low
fees in 2026~\cite{kumar2026clientdiversity}.

We measure it from the chain.
We follow every close of every public channel since Lightning's first channels in January~2018, find the force
closes of channels outside the public graph by the fingerprint of their commitment transaction, and follow each of
their \NOutputs{} outputs to the transaction that spent it: \NClosesAll{} closes in all, \NForceAll{} of them force
closes. Fee statistics for every block, from our own node, place each transaction in the fee market of its block.
The main answer is that fee spikes do not bring a wave of force closes. In the week of a spike and the week after,
force closes were \PlEvAllNow{} times their level in the four weeks before, while ordinary weeks of the same years
show \PlEvAllNowMed. That holds for pre-anchor channels, whose commitment fee is fixed when it is signed, as for anchor
channels, whose fee can be raised at broadcast. The cost of a force close lies elsewhere: in its ratio to a cooperative close, in
the days the closer's funds stay locked, and in the exposure of pending payments when fees are high. The wait
outweighs the fees only in low-fee years, and the exposure has rarely become a loss. We make five contributions.

\begin{itemize}
\item \textbf{A census with rates.} Force closes became rarer per channel, and private channels now produce most of
them. Of the public channels funded in 2019, \CohortOneYearNineteen\% were force-closed within a year; of those
funded in 2025, \CohortOneYearTwentyFive\%. The force share of public closes fell from \ForceShareTwentyThree\% in
2023 to \ForceShareTwentySix\% in 2026. In 2026 force closes of channels outside the public graph outnumbered public
ones \UnlOverPubTwentySix{} to one (\S\ref{sec:census}).
\item \textbf{The cost of a force close in fees, time and block space.} A force close costs
\CostRatioMin--\CostRatioMax{} times a cooperative close in fees and, since 2022, about three times its block space; the
closer waits a median of \ToLocalMedDays{} days for its own balance, which outweighs the fees only in the low-fee
years 2025--26. Counting the open as well, a public channel's whole life cost a median of \LcMed{}~sat on-chain. When
fees are high, HTLC timeouts confirm late, and anchor channels' timeouts less often than pre-anchor ones. Yet
forwarders lost few payments: \SplitResForwarderPayments{} since \SplitResFromYear, all with the preimage on-chain
first. The only losses in the order a replacement-cycling attack leaves were a test that one operator ran against its
own node, and 2026's one wave of penalties matches a splicing bug already fixed in Core Lightning
(\S\ref{sec:cost}).
\item \textbf{The anchor money trail.} Anchor outputs became third-party dust. Owners spent
\OwnerPctTwentyTwo\% of anchors in 2022 and \OwnerPctTwentySix\% in 2026; third parties sweep the rest, from 2025 at
the first block they can, and fees took \MinersShareTwentySix\% of the swept value in 2026. No P2WSH anchor was left
unspent, while no third party has swept a simple-taproot anchor, so the premise of an open LND proposal to stop
sweeping anchors holds only for the older kind (\S\ref{sec:anchors}).
\item \textbf{Fee regimes.} Fee spikes did not bring a wave of force closes, judged against placebo weeks of
the same years. Fixed commitment fees overpaid in calm
markets, and all of Lightning's closes and required spends took at most
\BillFeePctMax\% of Bitcoin's transaction fees in any year (\S\ref{sec:fees}).
\item \textbf{The first zero-fee commitments.} The first zero-fee commitment closed on \ZfFirst, before the
specification was merged. Across the anchor force closes we observe, the new format would have saved
\CfSavingPct\% of the anchor mechanism's block space (\S\ref{sec:zerofee}).
\end{itemize}

This is the third paper of a programme on the Lightning Network's resilience. The first asked whether client diversity
protects the network when one implementation fails~\cite{kumar2026clientdiversity}, the second who patches
it~\cite{kumar2026upgradedynamics}, and this one what a failed channel costs on-chain. All data behind the figures,
tables and in-text numbers are released with the code. We report aggregates only: no node, address or transaction is
named.

\section{Background and Related Work}
\label{sec:background}

\subsection{Force closes}
A channel is a 2-of-2 multisignature output on the chain. Each party holds its own version of the latest commitment
transaction, which spends that output to the two balances and to one output per pending payment (HTLC). In the
holder's version, the holder's own balance (\texttt{to\_local}) is locked for \texttt{to\_self\_delay} blocks, a value
the peer chose at opening, while the peer's balance (\texttt{to\_remote}) is not~\cite{bolt3}. The delay gives the peer
time to react if the holder broadcasts an old, revoked commitment: the peer can then take every output with a penalty
(justice) transaction~\cite{poon2016bitcoin}. A force close is the broadcast of a commitment; the broadcaster waits out
the delay before sweeping its balance.

Pending payments add deadlines. An offered HTLC returns to the sender after its expiry height
(\texttt{cltv\_expiry}) through a pre-signed HTLC-timeout transaction, or, on the peer's commitment, directly; a
received HTLC is claimed with the payment's preimage through an HTLC-success transaction~\cite{bolt3,bolt5}. A node
that forwards a payment holds two HTLCs for it, and the incoming one expires a margin later than the outgoing one,
the \texttt{cltv\_expiry\_delta} the node advertises in its channel updates~\cite{bolt7}. If the node cannot time out
its outgoing HTLC on-chain within that margin, the downstream node can claim it with the preimage while the
upstream one times out the incoming HTLC, and the forwarder pays for a payment it is not paid for. Flood-and-loot,
time-dilation and replacement-cycling attacks all work by making such a transaction confirm too
late~\cite{harris2020flood,riard2020timedilation,riard2023replacement}.

\subsection{Who pays the fee}
Commitment transactions are signed when the channel state changes and broadcast later, at a fee rate nobody knew at
signing. Three formats answer this differently.

\emph{Pre-anchor commitments} carry a fee fixed at signing. The channel's opener proposes it with
\texttt{update\_fee}, and a peer that finds the rate unreasonable can fail the channel~\cite{bolt2}. The broadcaster
cannot raise the fee later, because its own output is time-locked. These disagreements are a commonly cited cause of
force closes when fees move fast.

\emph{Anchor commitments} (specified in 2020~\cite{bolts688anchors}) add two 330-sat anchor outputs, one per party,
which their owner can spend at once with a child transaction that pays for both (child pays for parent, CPFP). Sixteen
blocks after the commitment confirms, anyone can spend either anchor, so that unused anchors do not stay in the UTXO
set~\cite{bolt3}. The peer's \texttt{to\_remote} output carries a one-block delay, so that the anchors are the only
outputs a child can spend at once, and the CPFP carve-out, a Bitcoin Core mempool rule, let each party attach one
child even if the other had filled the package limits~\cite{core31}. Second-stage HTLC transactions are signed so that
the holder can add inputs to pay their fee~\cite{bolt3}. Simple taproot channels follow the same design with taproot
outputs and anchor scripts of their own~\cite{bolts995taproot}.

\emph{Zero-fee commitments}~\cite{bolts1228zerofee} pay no fee at all. The commitment is a version-3 (TRUC)
transaction~\cite{bip431} with one keyless pay-to-anchor (P2A) output in place of the two anchors, and it is relayed
together with a child that pays for both, which Bitcoin Core supports since version 28.0~\cite{core28}. TRUC's topology limits replace
the carve-out as the defence against transaction pinning~\cite{optechpinning}, and Bitcoin Core 31.0 removed the
carve-out~\cite{core31}. The specification merged the format in May~2026; channels move to it as operators upgrade
their software, which in the Lightning Network takes months to years~\cite{kumar2026upgradedynamics}.

\subsection{Related work}
BitMEX Research first counted non-cooperative closes from the chain, private channels included, and found more than
60,000 by January~2020~\cite{bitmex2020closures}. Zabka et al.\ classified how public channels close
\cite{zabka2022empirical}; Grötschla et al.\ followed public channels from funding to the spending of their closing
outputs~\cite{grotschla2025lifecycle}; Antonelli et al.\ predict which channels will close and
how~\cite{antonelli2026closures}. Kappos et al.\ and Romiti et al.\ used closing transactions to link channels to
on-chain wallets~\cite{kappos2021empirical,romiti2021crosslayer}. Our previous paper measured force closes around
the 2026 Core Lightning shutdown~\cite{kumar2026clientdiversity}. None of these accounts for what a force close costs
or follows the anchor outputs.

The fee-bumping design has been argued from small samples. Daftuar analysed the 2023 transactions that spent
anchor-like outputs, to test whether Lightning's commitment spends would fit version-3 relay
rules~\cite{daftuar2024v3}. An open LND issue proposes to stop sweeping ``dangling'' anchors, on the argument that
third parties sweep them more efficiently, and gives the break-even fee rate without
data~\cite{lnd8997}. The security literature models what happens when closing transactions confirm late: mass closes
that congest the chain~\cite{harris2020flood,sguanci2023massexit}, nodes kept from seeing new
blocks~\cite{riard2020timedilation}, pinned or cycled
replacements~\cite{optechpinning,riard2023replacement}, HTLC slots jammed by attackers~\cite{mizrahi2021congestion}
and fees taken from revoked commitments~\cite{harding2024irrevocable}. We measure how often, and in what fee
conditions, such deadlines are missed in practice.

\section{Data}
\label{sec:data}

\paragraph{The chain}
We scan every block from block \ScanStartHeight{}, before Lightning's first channels, to block \CutoffHeight{}, the last
block mined before 1~November~2026 00:00~UTC, with our own Bitcoin Core node. For every block we keep its fee
statistics (\texttt{getblockstats}: the size-weighted percentiles of fee rates, total fees and total weight). A full
index of the chain built from the same node, with every transaction's locktime and every input's witness, gives the
HTLC expiries and the funding fees used in \S\ref{sec:htlc} and \S\ref{sec:lifecycle}.
On the blocks from 2021, which we also queried from the node directly, its fee totals agree with
\texttt{getblockstats} on every block.

\paragraph{Public channels}
Our list of public channels joins our own gossip since May~2023, mempool.space's record of closed channels and their
close reasons~\cite{mempoolspace}, and the lnresearch gossip archive for the earlier
years~\cite{lnresearchtopology}. A short channel ID encodes the block, transaction and output of the channel's funding
output, so the list resolves to \NFundingOutpoints{} funding outpoints, each to the transaction that spends it: the
close. The list holds few of the channels that closed in 2018: the gossip archive and mempool.space's record become
dense only in January~2019, and only about 1\% of 2018's force closes are of listed channels. For 2018 we therefore
report force closes only, listed or not, which the commitment fingerprint finds completely; public, cooperative,
force-share and per-channel series start in 2019, and January~2019 is still incomplete.

\paragraph{Closes and their outputs}
A close is cooperative or forced; a close that funds a new channel in the same transaction is a splice and is left
out. Force closes of channels that are not in the list are found by the fingerprint of the commitment transaction
alone (\S\ref{sec:methods}); we call them \emph{unlisted}. Up to the cutoff the data hold \NPublicCloses{} closes of
public channels (\NPublicMutual{} cooperative, \NPublicForce{} forced) and \NUnlistedForce{} unlisted force closes.
For each of the \NOutputs{} outputs of these \NForceAll{} force closes we record the transaction that spent it, if one
had by the cutoff (\NOutputsSpent{} did), and classify the spend from its witness script: the broadcaster's delayed
sweep, the peer's sweep, an HTLC transaction or claim, an anchor spend by its owner or by a third party, or a penalty.
Statistics about what happens to a close's outputs use only closes confirmed at least 4,032 blocks (four weeks) before
the cutoff, so that their outputs had time to be spent.

\paragraph{Validation}
We check the closes two ways. First, the chain index above is a separate extraction of the chain, by a different
program: in it, the transaction that spends each public channel's funding output is the close we recorded, for all
\ValLakeCompared{} closes that could be compared, and its shape, with or without the commitment fingerprint, matches
our force or cooperative label in every case. Second, mempool.space gives a close reason for
\ValCoverageYearMinPct--\ValCoverageYearMaxPct\% of the public channels that closed in each year from 2019 to 2025,
and none for 2026. It labels closes from the chain as well, so agreement with it is a check
of consistency, not independent validation. Where it gives a reason, for \ValCompared{} of our public closes
(\ValCoveragePct\%), our label agrees on \ValAgreePct\%.

\paragraph{Client labels}
For the appendix we attach to each public close the implementations at its two ends, using the labels and
classifier of our first paper on the day of the close~\cite{kumar2026clientdiversity}. They exist from May~2023 and
inherit that paper's accuracy.

\paragraph{What the chain does not show}
The chain does not show which party broadcast a force close, so every cost is per channel, both parties together. It
shows fee bumps that confirmed, not those that were tried and failed. Private channels appear only when they force
close: their cooperative closes look like ordinary payments, so their number and their force-close rate are unknown.
\S\ref{sec:limitations} lists the consequences.

\section{Methods}
\label{sec:methods}

\subsection{Force closes and their formats}
A commitment transaction carries a fingerprint: the upper byte of its locktime is \texttt{0x20} and that of its single
input's sequence is \texttt{0x80}, because the two fields encode the obscured commitment number~\cite{bolt3}. We call a
transaction a force close when it has this fingerprint and spends a 2-of-2 multisignature output or a taproot output by
its key path. Its format is \emph{zero-fee} if it is a version-3 transaction with a pay-to-anchor output,
\emph{simple taproot} if it spends a taproot output, \emph{anchor} if it has at least one anchor output, and
\emph{pre-anchor} otherwise.

\subsection{Rates and cohorts}
Counts of closes follow the size of the network, so we also report rates. A public channel is \emph{open} at a height
when its funding output is still unspent. Channels whose nodes have left the network count as open until they are
closed on-chain, which is right for a rate of on-chain closes but inflates the denominator as such channels
accumulate. The force-close rate is the number of public force closes per 1,000 open public channels per month (or
week). Following our first paper~\cite{kumar2026clientdiversity}, a \emph{mass-close event} is a node party to at least
20 public force closes on one UTC day; \emph{organic} force closes are the others. The \emph{cohort incidence} at age
$a$ is the share of the public channels funded in a year that were force-closed by age $a$, among those funded at
least $a$ before the cutoff. Open channels do not enter it, so it is not biased by abandoned channels; we draw a curve
while at least half of its cohort is observed.

\subsection{The cost of a force close}
We count every transaction a force close requires, in three currencies. \emph{Fees}: the commitment's fee; an anchor
spend by its owner in the commitment's block, a CPFP fee bump, counted in full; HTLC transactions and claims; the
broadcaster's delayed \texttt{to\_local} sweep; the \texttt{to\_remote} sweep of anchor channels, which the one-block
delay makes a separate step; and later anchor spends by their owner, which we call cleanup. A spend that batches several
inputs is split equally among them; charging each close the whole transaction gives an upper bound. Not counted: the
\texttt{to\_remote} spend of a pre-anchor channel, a plain wallet output like those of a cooperative close; third-party
anchor sweeps, which are paid from the anchors' own 330~sat; and penalties, reported apart. Each year's median is set
against the median fee of a cooperative close in the same year. \emph{Time}: the days from the close until the
broadcaster's \texttt{to\_local} can be swept, from the \texttt{to\_self\_delay} revealed in its witness, and the
balance times that wait (BTC-days). To set the wait against the fees without assuming a cost of capital, the
break-even rate is the yearly return on the locked balances at which a year's BTC-days would cost as much as that
year's force-close fees. \emph{Block space}: the virtual bytes of the same transactions.

\subsection{HTLC deadlines}
\label{sec:htlc-method}
The chain gives each HTLC's expiry: an HTLC-timeout transaction's locktime is the expiry, because its pre-signed
signature commits to it, and a received HTLC's script, revealed when it is spent, contains it~\cite{bolt3}. We measure
how many blocks after its expiry each HTLC-timeout transaction confirmed and, for HTLC-success transactions, how many
blocks before. The yardstick is the margin a forwarding node keeps, its \texttt{cltv\_expiry\_delta}: we use
\HtlcMarginTight{} blocks, the tenth percentile of the values public channels advertised in our daily gossip
snapshots from \CltvDeltaFrom{} to the cutoff, and \HtlcMarginMid, their median.
The fee condition is the median fee rate of the block at the expiry. To compare formats we use the HTLCs whose
commitment had confirmed by the expiry, so that the delay is the HTLC transaction's own. Both HTLC scripts also contain
a hash of the payment hash; until point time-locked contracts replace hashes, every hop of a payment and every part of
a multi-part payment share it, so the same hash in two force closes marks one payment.

\subsection{Lifecycle cost}
For each public channel closed at least four weeks before the cutoff, the lifecycle cost adds the channel's share of
its funding transaction's fee to the cost of its close and required spends (as above, or the fee of the cooperative
close). A funding transaction's fee is split equally among the channels it opens; its other outputs count as the
opener's change and pay nothing, because the change exists only because of the open. A splice is charged once, to the
new channel's open. Rules that also charge change outputs, by value or by size, are reported as a robustness check.

\subsection{Anchors and who sweeps them}
Each P2WSH anchor is spent by its owner in the commitment's block (a bump), by its owner later (cleanup), by a third
party (an input with an empty signature, possible from the sixteenth block), or not at all. For a same-block owner
spend we also test whether the child raised the fee rate of the package, which separates a bump from cleanup that
happened to confirm in the same block. Third-party sweeps are collected chain-wide, including anchors of commitments we
never saw, and each sweep's 330~sat per anchor is split between the fee (to miners) and the outputs (to the sweeper).
The break-even fee rate, at which the fee for one anchor input equals the anchor's value, is \BreakEven~sat/vB. To
measure concentration we link sweep transactions that spend each other's outputs or pay the same address; each
connected group is an operator. Sweeps whose outputs are all \texttt{OP\_RETURN} give everything to miners and are
counted apart. We report shares only and name no operator or address.

\subsection{Fee spikes}
\label{sec:spike-method}
To test whether fee spikes bring force closes we use an event study and a regression. A \emph{spike week} has a median
block fee rate at least \SpikeRule{} times that of the previous eight weeks and at least \SpikeMinFee~sat/vB; only the
first week of a cluster counts. We compare organic public force closes in the spike week and the week after with the
four weeks before, pooled over all spikes. The regression, by week, relates the logarithm of organic public force
closes per open channel to the logarithm of the median block fee rate, with year fixed effects and a linear trend, so
that weeks are compared within the same fee regime rather than 2023 with 2025. We fit it with the fee rate of the same
week, of one to four weeks earlier and of one week later.

We judge both against placebos that keep the force-close series as observed and break only its timing against fees.
(A) For each spike, an ordinary week of the same year is drawn at random, away from every real spike
(\PlWeekDraws{} draws). (B) Each year's fee series is cut into \PlBlockWeeks-week blocks, which are shuffled within the
year (\PlBlockDraws{} draws). (C) The fee series' deviations from its year means and trend are rotated by every shift of
at least \PlMinShift{} weeks (\PlShifts{} shifts), which keeps all of its autocorrelation. A result's percentile is the
share of draws below it. The random seed (\PlSeed) and the verdict rules were fixed before the full run: a wave counts
only if the spike ratio is above 95\% of placebo~A, and a delayed rise only if the largest coefficient of lags one to
four beats the largest in each draw at $p<0.05$ under both B and C. We do not use Newey--West intervals: under the
placebos, nominal 5\% tests rejected in \PlNwRejectBlockMin--\PlNwRejectBlockMax\% of draws of B and
\PlNwRejectCircMin--\PlNwRejectCircMax\% of draws of C.

\emph{Fee guessing.} A commitment's fee rate is compared with the median and the tenth percentile of the fee rates in
its own block. \emph{Sweep timing.} A third-party sweep's block is compared with the block sixteen blocks after the
close, when the sweep first became possible. \emph{Closing bill.} The fees and block space of every close and required
spend, public or not, plus third-party sweeps, are compared with all transaction fees and block weight in each year.

\subsection{Zero-fee counterfactual}
For every anchor force close we compare the block space of the anchor mechanism, two 43-vB anchor outputs and the
transactions that spend them, with that of a zero-fee commitment: one 13-vB pay-to-anchor output, spent in the same
package, and a child with a wallet input and change for the closes that were not already bumped. An owner's anchor
input takes 69.75~vB, a third party's 51.75~vB (measured on mainnet sweeps) and a pay-to-anchor input 41.25~vB.

\section{How Often Channels Force-Close}
\label{sec:census}

\begin{table*}[t]
\centering
\caption{Lightning channel closes on the chain, 2018--2026. All force: every force close, public or not. Public: channels in our list of every public channel. Unlisted: force closes of channels outside it, found by the commitment pattern.}
\label{tab:census}
\footnotesize
\begin{tabular}{lrrrrrrrrrr}
\toprule
Year & All force & Coop. & Force & Force \% & Force BTC & Age (d) & Unlisted & Unl. BTC & Pub. legacy \% & Unl. taproot \% \\
\midrule
2018\textsuperscript{b} & 35{,}218 & -- & -- & -- & -- & -- & -- & -- & -- & -- \\
2019\textsuperscript{c} & 52{,}148 & 30{,}404 & 30{,}001 & 49.7 & 679 & 80 & 22{,}147 & 411 & 100.0 & 0.0 \\
2020 & 29{,}877 & 28{,}785 & 17{,}813 & 38.2 & 581 & 159 & 12{,}064 & 198 & 100.0 & 0.0 \\
2021 & 39{,}539 & 64{,}048 & 27{,}393 & 30.0 & 1{,}123 & 94 & 12{,}146 & 179 & 87.5 & 0.0 \\
2022 & 73{,}569 & 92{,}645 & 57{,}896 & 38.5 & 3{,}226 & 133 & 15{,}673 & 264 & 44.8 & 0.0 \\
2023 & 76{,}869 & 62{,}481 & 49{,}998 & 44.5 & 3{,}363 & 151 & 26{,}871 & 764 & 40.2 & 0.1 \\
2024 & 76{,}634 & 50{,}874 & 28{,}888 & 36.2 & 2{,}208 & 215 & 47{,}746 & 1{,}410 & 33.9 & 0.6 \\
2025 & 71{,}498 & 58{,}137 & 21{,}518 & 27.0 & 1{,}975 & 242 & 49{,}980 & 838 & 23.4 & 3.2 \\
2026\textsuperscript{a} & 68{,}463 & 62{,}311 & 15{,}440 & 19.9 & 2{,}279 & 142 & 53{,}023 & 974 & 25.6 & 23.1 \\
\bottomrule
\end{tabular}
\par\smallskip\parbox{\linewidth}{\scriptsize \textsuperscript{a}January--October. \textsuperscript{b}From 12 January (block 503,816). Our list of public channels holds few of the channels that closed in 2018, so only the force-close total is shown. \textsuperscript{c}The list's coverage is still ramping up in January 2019. Age: median age at force close. Pub.\ legacy: share of public force closes with a pre-anchor commitment. Unl.\ taproot: share of unlisted force closes of simple taproot channels. Splices (59) are not shown.}
\end{table*}

\begin{figure*}[t]
\centering
\includegraphics[width=\textwidth]{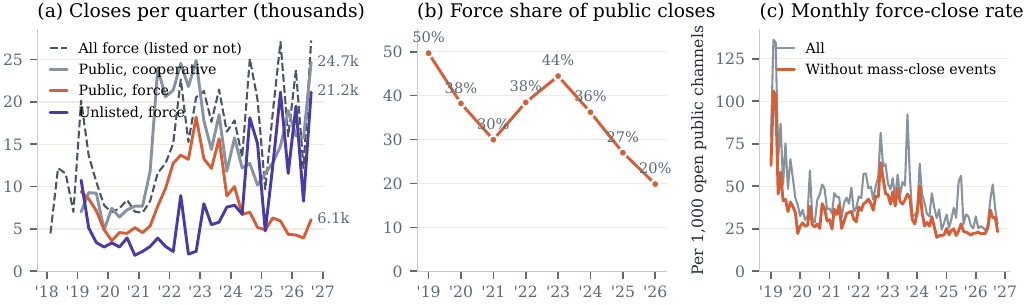}
\caption{Lightning channel closes on the chain. (a) Closes per quarter: all force closes, listed or not (dashed,
from January~2018), and from 2019 the cooperative and force closes of public channels and the force closes of
channels outside our list of public channels (unlisted). (b) Force share of public closes by year, from 2019.
(c) Public force closes per 1,000 open public channels per month, from 2019, with and without mass-close events
(one node party to at least \MassMin{} force closes on one UTC day).}
\label{fig:census}
\end{figure*}

\paragraph{Force closes became rarer, but less than their counts suggest}
Public force closes peaked at \PubForceTwentyTwo{} in 2022 and fell to \PubForceTwentySix{} in the first ten months of
2026 (Table~\ref{tab:census}). Part of that fall is a smaller network: open public channels peaked at \OpenPeak{} and
stood at \OpenAtCutoff{} at the cutoff. Part is more cooperative closing: cooperative closes released
\MutualBtcTwentyFour~BTC of capacity in 2024 and \MutualBtcTwentyFive~BTC in 2025. The force share of public closes
therefore fell faster than the risk to a channel. It went from \ForceShareTwentyThree\% in 2023 to
\ForceShareTwentySix\% in 2026, while the organic force-close rate went from \OrganicRateTwentyThree{} to
\OrganicRateTwentySix{} per 1,000 open channels a month (Fig.~\ref{fig:census}c).

\paragraph{2019 reads highest, but our list may be incomplete}
In 2019 the force share (\ForceShareNineteen\%) and the organic rate (\OrganicRateNineteen{} per 1,000 open channels a month)
were the highest of any year. We do not read this as a peak, because our channel list may still have been filling in. In
February and March~2019 the organic rate was \RateOrganicFebNineteen{} and \RateOrganicMarNineteen{} per 1,000, and
no later month exceeded \RateOrganicMaxLater.

\begin{figure}[t]
\centering
\includegraphics[width=\columnwidth]{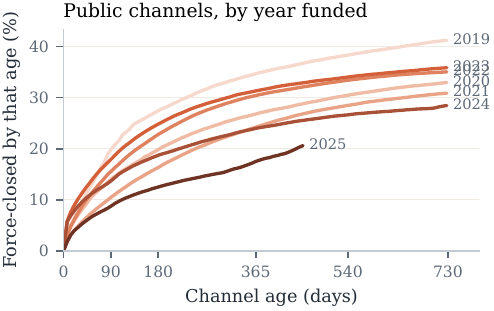}
\caption{Share of public channels force-closed by each channel age, by the year the channel was funded (2019--2025).
A curve ends where less than half of its cohort has been observed for that long.}
\label{fig:cohorts}
\end{figure}

\paragraph{The risk to a new channel fell}
\CohortOneYearNineteen\% of the public channels funded in 2019 were force-closed within a year, against
\CohortOneYearTwentyFive\% of those funded in 2025 (Fig.~\ref{fig:cohorts}). The fall was not steady: the share
dropped to \CohortOneYearTwentyOne\% for the 2021 cohort, rose to \CohortOneYearTwentyThree\% for 2023's and fell
again after. Cohort incidence does not depend on the count of open channels, which abandoned channels inflate, so it
is the cleaner measure of the trend.

\paragraph{Much force closing is one node closing many channels at once}
Every year has \MassEventsMin--\MassEventsMax{} mass-close events, which hold \MassShareMin--\MassShareMax\% of public
force closes; the largest put \MassLargest{} channels on-chain in one day. Without them the rate still falls, so the
decline is not one operator's. The 2026 peak is such an episode. In August~2026 the rate reached \RateAugTwentySix{}
per 1,000 open channels, \RateOrganicAugTwentySix{} without mass-close events, while open channels fell from
\OpenJulTwentySix{} in July to \OpenAtCutoff{} at the cutoff. That month brought the Core Lightning shutdown and the
largest contraction of the public graph. Our first paper found the contraction driven by large operators leaving,
and organic force closes of channels with a Core Lightning node above their usual rate after the
shutdown~\cite{kumar2026clientdiversity}.

\paragraph{Most force closes now happen outside the public network}
In 2026, unlisted force closes outnumbered public ones \UnlOverPubTwentySix{} to one. From 2024 on they are almost all
private channels; before 2024 they include public channels missing from our list (\S\ref{sec:limitations}).
Private channels are often those of mobile wallets~\cite{bitmex2020closures}, and \UnlTaprootPctTwentySix\% of
2026's unlisted force closes were simple taproot channels. The largest burst in the data is unlisted: on 1~September~2024, \SepEventCloses{} unlisted force closes
holding \SepEventBtc~BTC confirmed within about \SepEventHours~hours (\SepEventBlocks{} blocks), \SepEventLegacy{} of
them pre-anchor.

\paragraph{Most of 2026's pre-anchor closes are one cluster}
Pre-anchor commitments were \PubLegacyPctTwentyOne\% of public force closes in 2021 and \PubLegacyPctTwentySix\% in
2026 (Table~\ref{tab:census}), but most of the 2026 share is one cluster. Five nodes were party to
\LegacyTopFivePct\% of 2026's pre-anchor public force closes, four of them LDK nodes that announced no anchor
support~\cite{kumar2026clientdiversity}, and their channels lasted a median of \LegacyClusterAgeDays~days. Without
closes involving those nodes, pre-anchor commitments were \LegacyPctExClusterTwentySix\% of 2026 public force closes,
and those channels were old (median \LegacyOldAgeYears~years).

\paragraph{Client pairs}
Among client pairs with more than 1,000 closes since May~2023, the force share ranges from \PairForcePctLndLnd\%
(LND--LND) to \PairForcePctClnCln\% (CLN--CLN), with one outlier that a single node explains
(Appendix~\ref{app:pairs}; labels from~\cite{kumar2026clientdiversity}).

\section{What a Force Close Costs}
\label{sec:cost}

\begin{table*}[t]
\centering
\caption{What a public force close costs, against a cooperative close, 2019--2026 (medians, sat unless stated).}
\label{tab:costs}
\footnotesize
\begin{tabular}{lrrrrrrrrrr}
\toprule
Year & Force closes & Commit. & Total & p90 & Coop. & Ratio & vB & Coop.\ vB & >1\% cap. & Days \\
\midrule
2019 & 30{,}001 & 4{,}900 & 7{,}460 & 24{,}802 & 3{,}485 & 2.1 & 270 & 138 & 54.9 & 1.0 \\
2020 & 17{,}813 & 9{,}051 & 14{,}318 & 36{,}545 & 3{,}994 & 3.6 & 271 & 168 & 63.1 & 1.0 \\
2021 & 27{,}393 & 2{,}583 & 3{,}984 & 31{,}397 & 697 & 5.7 & 302 & 168 & 41.0 & 1.0 \\
2022 & 57{,}896 & 2{,}021 & 3{,}342 & 7{,}920 & 551 & 6.1 & 521 & 169 & 16.1 & 1.7 \\
2023 & 49{,}998 & 2{,}872 & 7{,}346 & 45{,}588 & 2{,}005 & 3.7 & 595 & 192 & 32.1 & 2.1 \\
2024 & 28{,}888 & 3{,}012 & 6{,}332 & 36{,}635 & 2{,}124 & 3.0 & 581 & 192 & 29.6 & 2.0 \\
2025 & 21{,}518 & 730 & 1{,}292 & 3{,}912 & 262 & 4.9 & 598 & 192 & 11.7 & 1.7 \\
2026\textsuperscript{a} & 14{,}435 & 597 & 925 & 2{,}363 & 197 & 4.7 & 585 & 192 & 7.5 & 2.5 \\
\bottomrule
\end{tabular}
\par\smallskip\parbox{\linewidth}{\scriptsize \textsuperscript{a}Closes to 4 October 2026, so that outputs had four weeks to be spent. Total: every fee the force close requires (commitment, fee bump, HTLC resolution, sweeps, anchor cleanup); batched spends split per input. Ratio: median total over the median cooperative close fee. vB: block space, the same way. $>$1\% cap.: share of force closes whose fees exceed 1\% of channel capacity. Days: median wait before the broadcaster's own balance can be swept.}
\end{table*}

\begin{figure*}[t]
\centering
\includegraphics[width=\textwidth]{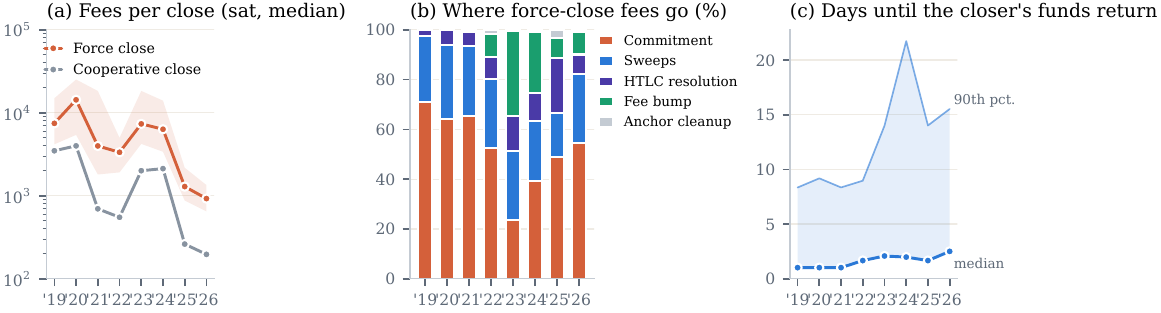}
\caption{The cost of a public force close. (a) Median fees per close, force and cooperative (log scale; band:
interquartile range of force closes). (b) Where force-close fees go: the commitment, sweeps of the broadcaster's
balance and of anchor channels' \texttt{to\_remote}, HTLC resolution, fee bumps (owner CPFP children in the
commitment's block) and later anchor cleanup. (c) Days until the broadcaster's own balance can be swept: median and
90th percentile.}
\label{fig:costs}
\end{figure*}

\subsection{Fees, time and block space}

\paragraph{Cheap in fees, a multiple of a cooperative close}
In fees alone a force close is cheap: the median public force close cost \CostMedTwentySix~sat in 2026
(Table~\ref{tab:costs}). The median was highest in 2020, at \CostMedTwenty~sat, against \CostMedTwentyThree~sat in
2023. Against a cooperative close of the same year it is dear: \CostRatioMin--\CostRatioMax{} times the median
cooperative close in every year. The fee market sets the level, not the ratio. Force closes paid
\CostTotalBtcAll~BTC in fees over the whole period, \CostTotalBtcTwentyThree~BTC of it in 2023, more than in any other
year.

\paragraph{When fees are high, the money goes to bumps and HTLCs}
In 2023, fee bumps took \BumpShareTwentyThree\% of all force-close fees and HTLC resolution \HtlcShareTwentyThree\%
(Fig.~\ref{fig:costs}b). That year \AnchorBumpedPctTwentyThree\% of anchor force closes were bumped by a child in the
commitment's block, and \BumpPctHtlcTwentyThree\% of those with pending HTLCs. In the low-fee years the commitment
and the sweeps make up most of the cost.

\paragraph{Anchors doubled the block space}
An anchor force close uses about twice the block space of a pre-anchor one (median \AnchorVbMed{} against
\LegacyVbMed~vB): two extra outputs, a \texttt{to\_remote} output that needs a sweep of its own, and the anchors'
spends. Since 2022 a force close has taken \VbRatioMin--\VbRatioMax{} times the block space of a cooperative close.
Block space does not depend on the fee regime, which makes it the currency in which formats compare most directly
(\S\ref{sec:zerofee}).

\paragraph{The closer waits days}
The broadcaster's own balance returns a median \ToLocalMedDays{} days after the close, and in the slowest tenth of
closes up to \ToLocalPNinetyMax{} days (Fig.~\ref{fig:costs}c). The wait is the \texttt{to\_self\_delay} the peer chose
at opening: \DelayOneDayPct\% of values are 144 blocks (one day) and \DelayTwoWeeksPct\% about two weeks. Over the
period, closers' balances were locked for \BtcDaysLockedAll{} BTC-days.

\paragraph{The wait outweighs the fees only when fees are low}
The break-even rate was \BreakEvenRateTwentyFive\% in 2025 and \BreakEvenRateTwentySix\% in 2026, so in the low-fee
years even a modest return on the locked funds made the wait the larger cost. Before 2025 it was
\BreakEvenRateEarlyMin--\BreakEvenRateEarlyMax\% a year. Over the whole period the wait cost more than the fees only
for funds earning more than \BreakEvenRateAll\% a year.

\paragraph{Relative to capacity, the cost fell}
In 2021, \OverOnePctCapTwentyOne\% of public force closes paid more than 1\% of the channel's capacity in fees; in
2026, \OverOnePctCapTwentySix\%.

\subsection{Penalties and a breach wave}
\label{sec:breach}

\paragraph{Penalties are rare}
Over the period, \PenaltyTxsAll{} penalty transactions took \PenaltyBtcAll~BTC from parties that broadcast a revoked
commitment, in public or private channels. Until September~2026 a typical month had \PenaltyMonthlyMed{} and none
had more than \PenaltyMonthlyMaxBefore{} (\PenaltyMonthlyMaxBeforeMonth). Only punished breaches are visible: a breach
that went unpunished looks like an ordinary force close.

\paragraph{A one-week breach wave}
Between \PenaltyWaveFirst{} and \PenaltyWaveLast, revoked commitments of \BreachWavePenalised{} private anchor
channels were broadcast, and every one was penalized within \PenaltyWaveMaxBlocks{} blocks. \BreachWaveChannels{} of
them (\BreachWaveBtc~BTC) were splices of channels opened a median of \BreachOpenToSpliceMed{} blocks earlier by one
funder, which also collected the penalties. The counterparty force-closed about \BreachSpliceToCloseMed{} blocks after
each splice; \BreachLookAlikesWeek{} more such closes that week used the current state. The remaining channel is an
unrelated breach of an older channel. The transactions on both sides follow Core Lightning's construction rules, and
the wave matches a bug that Core Lightning fixed in version 26.06.8, released on 19~September~2026~\cite{cln26068}. In
affected versions, a unilateral close after a splice had locked broadcast the commitment from the moment of the lock,
which is revoked as soon as the channel is updated again. Core Lightning locks a splice at three confirmations, which
is when the closes came. The bug explains why the old states were always post-splice states and why some closes used
the current state. The chain does not show what made each counterparty close at the lock, or whether the funder
caused it. This one week holds most of 2026's \PenaltyTxsTwentySix{} penalties and is the only
event of its kind in our data, which go back to 2018.

\subsection{Exposure: HTLC deadlines}
\label{sec:htlc}

\begin{figure*}[t]
\centering
\includegraphics[width=0.85\textwidth]{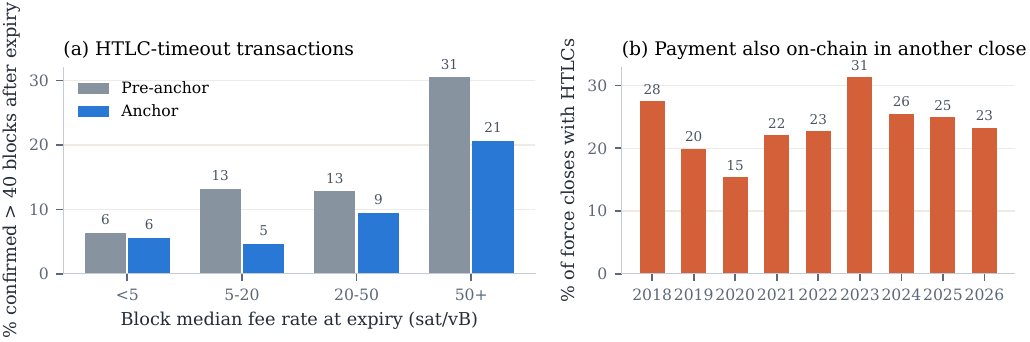}
\caption{HTLCs resolved on-chain. (a) Share of HTLC-timeout transactions confirmed more than \HtlcMarginTight{}
blocks after the HTLC's expiry, by the median fee rate of the block at expiry, for pre-anchor and anchor channels
whose commitment had confirmed by the expiry. (b) Share of force closes with an HTLC resolved on-chain whose payment
also appears in another force close, by year.}
\label{fig:htlc}
\end{figure*}

\paragraph{Most HTLC timeouts are prompt; the late tail grows with fees}
Of \HtlcTimeoutTxs{} HTLC-timeout transactions, the median confirmed \HtlcTimeoutMedMin--\HtlcTimeoutMedMax{} blocks
after its expiry in every year. The tail is what matters for a forwarding node. More than \HtlcMarginTight{} blocks
late were \HtlcLateEarlyMin--\HtlcLateEarlyMax\% of timeouts in each year from 2019 to 2022, then
\HtlcLateTwentyThree\% in 2023, the high-fee year, and \HtlcLateTwentyFour\% in 2024. In \HtlcCommitBeforeExpiryPct\%
of cases the commitment had confirmed before the expiry, so the delay was the HTLC transaction's own.

\paragraph{Anchors keep HTLC timeouts closer to the deadline when fees are high}
When the block median fee rate at expiry was 50~sat/vB or more, \HtlcLateSpikeLegacy\% of pre-anchor HTLC timeouts
confirmed more than \HtlcMarginTight{} blocks late, against \HtlcLateSpikeAnchor\% of anchor ones
(Fig.~\ref{fig:htlc}a). At 5--20~sat/vB the gap was \HtlcLateMidLegacy\% against \HtlcLateMidAnchor\%; below
5~sat/vB the two formats were alike. A pre-anchor HTLC transaction carries the fee fixed at signing, while an anchor
channel's can add inputs to pay a fee chosen at broadcast. This is the effect the anchor design was built for, and
unlike the commitment-level test of \S\ref{sec:fees} it shows up in the data.
The exception is 2024, when anchor channels' timeouts were late more often than pre-anchor ones
(\HtlcLateTwentyFourAnchor\% against \HtlcLateTwentyFourLegacy\%).

\paragraph{Successes mostly come early}
HTLC-success transactions confirmed a median \HtlcSuccessMarginMin--\HtlcSuccessMarginMax{} blocks before expiry in
every year. In most years only \HtlcSuccessLateMinOther--\HtlcSuccessLateMaxOther\% came at or after the expiry, when
the peer can already take the HTLC back. Three years had more: \HtlcSuccessLateTwentyFour\% in 2024,
\HtlcSuccessLateNineteen\% in 2019 and \HtlcSuccessLateTwentyThree\% in 2023.

\paragraph{One payment can take several channels on-chain}
The payment hash links one payment's HTLCs across closes (\S\ref{sec:htlc-method}). Of the force closes with an HTLC resolved on-chain, \CascadePct\% share a payment with another force close
(Fig.~\ref{fig:htlc}b): \CascadePayments{} payments, up to \CascadeMaxCloses{} closes for one payment, with a median
of \CascadeSpanMed{} blocks between the first and the last. \CascadeAllTimeoutPct\% of them were timed out on every
channel: a stuck payment, or the parts of one multi-part payment, taking several channels down along its route.

\paragraph{Split resolutions are rare losses}
From \SplitResFromYear, in \SplitResPayments{} payments a forwarding node's two HTLCs were resolved in opposite ways:
with the preimage on one channel and by timeout on the other, so one party lost the HTLC amount (\SplitResBtc~BTC in
all). In \SplitResForwarderPayments{} of them (\SplitResForwarderBtc~BTC) the upstream HTLC timed out while the
downstream one was paid, so the node between them paid out and was not paid; in \SplitResPreimageFirstPct\% of those
the preimage was already on-chain a median \SplitResGapMed{} blocks before the upstream timeout confirmed. The losses
do not cluster in the 2023 fee spike, which points at nodes that were offline or failed to act rather than at fee
competition. These counts leave out \SplitResStagedPairs{} pairs from 2026 (\SplitResStagedBtc~BTC) whose channels
used a \SplitResStagedToSelfDelay-block \texttt{to\_self\_delay}: a replacement-cycling test, described next, not
organic losses.

\paragraph{No loss has the order replacement cycling leaves}
In a replacement-cycling attack the downstream peer keeps replacing the forwarder's HTLC-timeout transaction in the
mempool, and lets its own preimage claim confirm only after the upstream HTLC has timed
out~\cite{riard2023replacement}. A successful attack therefore puts the preimage on-chain after the upstream timeout.
No forwarder loss from \SplitResFromYear{} has that order. The staged pairs do, and they are such an attack, run as a
test: one operator ran the upstream, forwarding and downstream nodes and cycled its own forwarder's HTLC-timeout
transactions on mainnet. Shared funding and direct transfers link the three nodes' wallets, and the transactions that
evicted the forwarder's timeouts carry a fingerprint we find nowhere else on the chain in that period. In each staged
pair the preimage confirmed no earlier than the upstream timeout, a median of \SplitResStagedPreimageAfterMed{} blocks
after. The test sees only
payments whose two HTLCs were both resolved on-chain, so it bounds attacks that succeeded that way, not attempts.

\subsection{The whole life of a channel}
\label{sec:lifecycle}

\begin{table*}[t]
\centering
\caption{What a public channel's life costs on-chain: funding share, close and required spends, by year of close (medians, sat unless stated).}
\label{tab:lifecycle}
\footnotesize
\begin{tabular}{lrrrrrrrrrrr}
\toprule
Closed & Channels & Force \% & Open & Coop. & Force & All & Range & p90 & \% cap. & Opens \% & BTC \\
\midrule
2019 & 60{,}405 & 49.7 & 2{,}429 & 8{,}322 & 11{,}021 & 9{,}581 & 7{,}842--9{,}581 & 28{,}253 & 1.29 & 36.3 & 8.00 \\
2020 & 46{,}598 & 38.2 & 2{,}362 & 9{,}061 & 19{,}174 & 12{,}750 & 10{,}708--12{,}750 & 38{,}299 & 1.51 & 36.6 & 8.55 \\
2021 & 91{,}441 & 30.0 & 925 & 2{,}127 & 8{,}329 & 3{,}368 & 2{,}701--3{,}368 & 30{,}606 & 0.39 & 47.0 & 10.11 \\
2022 & 150{,}542 & 38.5 & 361 & 1{,}162 & 4{,}249 & 2{,}153 & 1{,}790--2{,}153 & 7{,}859 & 0.10 & 44.9 & 6.04 \\
2023 & 112{,}481 & 44.5 & 1{,}069 & 3{,}864 & 10{,}119 & 6{,}106 & 5{,}090--6{,}106 & 33{,}969 & 0.22 & 30.0 & 20.51 \\
2024 & 79{,}763 & 36.2 & 2{,}290 & 5{,}544 & 10{,}710 & 7{,}032 & 5{,}437--7{,}032 & 31{,}803 & 0.18 & 44.3 & 14.07 \\
2025 & 79{,}656 & 27.0 & 477 & 907 & 2{,}374 & 1{,}261 & 937--1{,}261 & 5{,}331 & 0.03 & 62.0 & 2.40 \\
2026\textsuperscript{a} & 71{,}639 & 20.1 & 309 & 626 & 1{,}450 & 785 & 569--785 & 2{,}927 & 0.01 & 64.6 & 1.24 \\
\midrule
All & 692{,}525 & 35.8 & 723 & 1{,}970 & 6{,}645 & 3{,}435 & 2{,}804--3{,}435 & 23{,}171 & 0.15 & 39.7 & 70.92 \\
\bottomrule
\end{tabular}
\par\smallskip\parbox{\linewidth}{\scriptsize \textsuperscript{a}Closes to 4 October 2026 (four weeks for outputs to be spent). Open: the channel's share of its funding transaction's fee, split equally over the channel outputs of the transaction; other outputs are the funder's change and pay nothing. Range: the median lifecycle if every output, change included, pays for its own size, and with the equal split. Coop., Force: lifecycles ending in a cooperative or a force close; force closes as in Table~II. \% cap.: median lifecycle fees over capacity. Opens \%: opens' share of all lifecycle fees. BTC: all lifecycle fees. Closes from 2019: our list of public channels holds few of the channels that closed in 2018.}
\end{table*}

\begin{figure*}[t]
\centering
\includegraphics[width=0.85\textwidth]{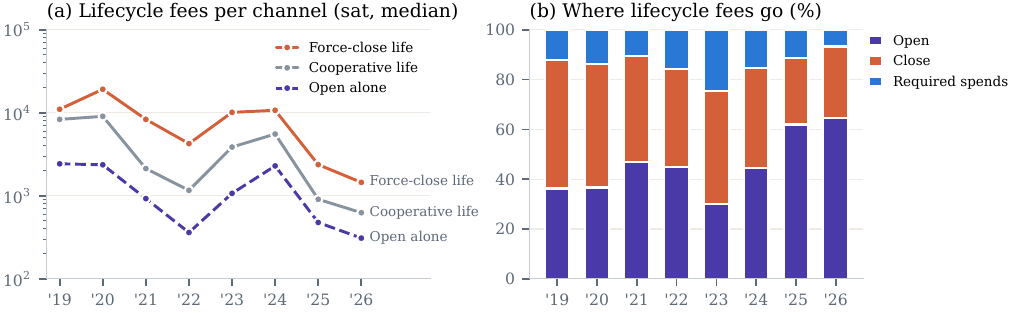}
\caption{A public channel's on-chain fees from funding to its last required spend. (a) Median by year of close, for
lives ending in a force close or a cooperative close, and the channel's share of its funding transaction alone (log
scale). (b) Where lifecycle fees go: the open, the close and the spends a force close requires.}
\label{fig:lifecycle}
\end{figure*}

\paragraph{A channel's life costs a few thousand satoshis}
Funding, closing and the spends a force close requires cost a public channel a median of \LcMed~sat
(Table~\ref{tab:lifecycle}): \LcOpenMed{} for its share of the funding transaction and \LcCloseMed{} for the close.
The median life cost \LcCapPct\% of the channel's capacity. The fee regime dominates. Channels closed in 2019--20 paid
the most, a median \LcMedEarlyRegime~sat; those closed in 2023--24 paid \LcMedHighRegime, and those closed in 2025--26
\LcMedLowRegime.

\paragraph{The open is a large part of the bill}
Opens paid \LcOpensPct\% of all lifecycle fees (\LcOpenBtc{} of \LcTotalBtc~BTC), and \LcOpensPctLowRegime\% for the
channels closed in 2025--26, which closed cheaply but had often opened in dearer blocks (Fig.~\ref{fig:lifecycle}b).

\paragraph{Counting the open narrows the gap between endings}
A life that ended cooperatively cost a median \LcCoopMed~sat, one that ended in a force close \LcForceMed{}, a ratio of
\LcForceCoopRatio{} (\LcForceCoopRatioMin--\LcForceCoopRatioMax{} by year), against \CostRatioMin--\CostRatioMax{}
for the close alone, because both kinds of channel paid the same to open. If change outputs also paid their share of
the funding fee, by value or by size, the median life would cost \LcRobustMinPct--\LcRobustMaxPct\% less; treating
funding transactions that also pay others as payment batches moves it by \LcRobustBatchPct\%. The lifecycles are
completed ones, so they favour short-lived channels (median life \LcLifeDaysMed{} days).

\section{What Anchor Outputs Are Used For}
\label{sec:anchors}

\begin{table*}[t]
\centering
\caption{What anchor outputs are used for. Left: anchors by the year of their force close. Right: third-party sweep transactions by the year they confirmed.}
\label{tab:anchors}
\footnotesize
\begin{tabular}{lrrrrrrrrrr}
\toprule
Year & Anchors & Owner \% & Third \% & Unspent \% & Bumped \% & Delay & Sweep txs & Per tx & sat/vB & Miners \% \\
\midrule
2021 & 6{,}674 & 93.7 & 6.3 & 0.0 & 8.3 & 15{,}054 & -- & -- & -- & -- \\
2022 & 62{,}826 & 81.3 & 18.7 & 0.0 & 12.5 & 78 & 1{,}341 & 9.1 & 1.00 & 23.9 \\
2023 & 70{,}131 & 35.2 & 64.8 & 0.0 & 43.8 & 1{,}708 & 1{,}117 & 37.3 & 3.44 & 65.3 \\
2024 & 68{,}631 & 15.3 & 84.7 & 0.0 & 18.3 & 3{,}308 & 3{,}021 & 21.7 & 4.90 & 84.1 \\
2025 & 102{,}663 & 11.9 & 88.1 & 0.0 & 9.0 & 16 & 9{,}474 & 9.5 & 2.50 & 82.6 \\
2026\textsuperscript{a} & 76{,}577 & 4.8 & 95.2 & 0.0 & 13.6 & 16 & 10{,}390 & 7.4 & 1.66 & 87.0 \\
\bottomrule
\end{tabular}
\par\smallskip\parbox{\linewidth}{\scriptsize \textsuperscript{a}Closes to 4 October 2026. Anchors: P2WSH anchors of all force closes, public or not. Bumped: share of public anchor force closes with an owner child in the commitment's block. Delay: median blocks from the close to a third-party sweep. Sweep txs: transactions spending anchors with an empty signature (chain-wide). sat/vB: their median fee rate. Miners: share of the anchors' value paid as fees. Break-even fee rate: 6.38 sat/vB.}
\end{table*}

\begin{figure*}[t]
\centering
\includegraphics[width=\textwidth]{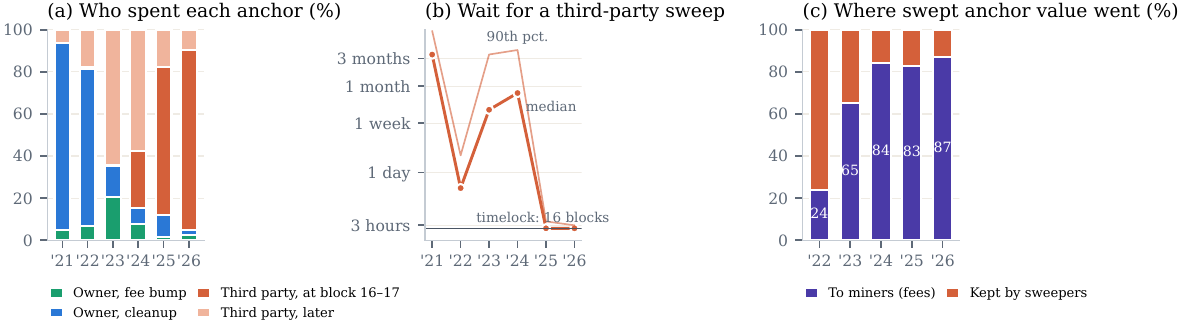}
\caption{The anchor money trail. (a) Who spent each P2WSH anchor, by year of the force close: its owner in the
commitment's block (fee bump) or later (cleanup), or a third party at blocks 16--17 or later. (b) Wait from the close
to a third-party sweep, public channels: median and 90th percentile (log scale). (c) Where swept anchor value went, by
year of the sweep: fees to miners, or kept by sweepers.}
\label{fig:anchors}
\end{figure*}

\paragraph{No P2WSH anchor is left unspent}
All \AnchorsAll{} P2WSH anchors created by force closes, public or not, at least four weeks before the cutoff had
been spent by the cutoff (Table~\ref{tab:anchors}). Anchors that their owners leave behind
do not stay in the UTXO set: third parties sweep them. That is the premise of an open LND proposal to stop sweeping
them~\cite{lnd8997}. For P2WSH anchors the data confirm it; simple-taproot anchors, below, are the exception.

\paragraph{Owners stopped using their anchors}
Owners spent \OwnerPctTwentyTwo\% of the anchors of 2022's force closes and \OwnerPctTwentySix\% of 2026's
(Fig.~\ref{fig:anchors}a, Table~\ref{tab:anchors}). Most owner spends
were cleanup after the commitment had confirmed, not fee bumps. Bumps peaked in 2023, when owners bumped
\BumpPctNoHtlcTwentyThree\% of anchor force closes without pending HTLCs and \BumpPctHtlcTwentyThree\% of those with
them. Even a same-block child is often not a bump: in 2026, \NotRaisedPctTwentySix\% of them did not raise the
package's fee rate, so they were cleanup that confirmed in the same block.

\paragraph{Sweepers wait for cheap blocks when fees are high}
Anyone may sweep an anchor from the sixteenth block after the close, but when fees were high nobody did so at once. The
median third-party sweep of a public channel's anchor came \ThirdMedDelayTwentyThree{} blocks after a 2023 close,
about \ThirdMedDelayDaysTwentyThree{} days, against exactly \ThirdMedDelayTwentyFive{} from 2025 (Fig.~\ref{fig:anchors}b). For 2023 closes,
\SweepCheaperPctTwentyThree\% of sweeps confirmed in a block cheaper than the one sixteen blocks after the close
(median fee rate \SweepPFiftyAtSweepTwentyThree{} against \SweepPFiftyAtSixteenTwentyThree~sat/vB). The anchors of
2021's closes waited for the cheap blocks of 2022.

\paragraph{Miners take most of the value}
Fees took \MinersShareTwentyTwo\% of the swept anchors' value in 2022, when sweeps paid a median
\SweepMedFrTwentyTwo~sat/vB, and \MinersShareLateMin--\MinersShareLateMax\% from 2024 (Fig.~\ref{fig:anchors}c). From 2022 to the cutoff,
third-party sweeps collected anchors worth \SweepValueBtc~BTC; \SweepMinersBtc~BTC of it went to miners and
\SweepSweepersBtc~BTC to the sweepers. A sweep pays only below \BreakEven~sat/vB, which is why sweepers
wait out high fees.

\paragraph{One operator collects most third-party anchors}
We find \SweepOperators{} sweeping operators from 2022 to the cutoff. Sweeps that give everything to miners, with only
\texttt{OP\_RETURN} outputs, took \SweepBurnPct\% of third-party anchors; of the rest, \SweepOperatorsOverOne{}
operators took more than 1\% each. The largest appeared on \SweepLargestFirst{} and took \SweepLargestPctTwentySix\%
of third-party anchors in 2026 and \SweepLargestPct\% of all since 2022, netting \SweepLargestNetBtc~BTC.

\paragraph{No third party has swept a simple-taproot anchor}
Simple taproot channels' anchors mostly stay unspent. Of \TaprootAnchors{} created from 2023 to the cutoff,
\TaprootAnchorsSpent{} were spent; of the \TaprootAnchorsTwentySix{} created in 2026, \TaprootAnchorsSpentTwentySix.
Every one of those spends is a key-path spend by the anchor's owner, and no input anywhere on the chain spends the
anchor's 16-block script path (\TaprootScriptPathChainwide{} inputs). That path needs the anchor's internal key,
which the commitment does not reveal; the specification has it revealed when the matching \texttt{to\_local} or
\texttt{to\_remote} output is swept~\cite{boltsimpletaproot}. The missing key does not explain the unspent anchors:
for \TaprootKeyExposedUnspent{} of the \TaprootAnchorsUnspent{} unspent anchors the key has been on-chain for a
median of \TaprootKeyExposedDaysMed{} days, and the sweepers that clear P2WSH anchors have not touched them. Why
they do not is a question for the implementations' developers.

\paragraph{The unswept taproot anchors are worth little to a sweeper}
The anchors whose key is on-chain hold \TaprootKeyPublicBtc~BTC in all, less than the largest P2WSH sweeper has
netted (\SweepLargestNetBtc~BTC). Owners' own spends stopped as well, even though adding the anchor to a sweep still
paid: in the first half of 2025, \TaprootCosweepToRemotePctEarly\% of owners' \texttt{to\_remote} sweeps also spent
the matching anchor, against \TaprootCosweepToRemotePctLate\% from July~2026 to the cutoff. The 2026 sweeps paid a
median \TaprootOwnerSweepFrLate~sat/vB, against \TaprootOwnerSweepFrEarly~sat/vB in early 2025, so including the
anchor was worth more to the owner in 2026, not less.

\section{Fee Regimes}
\label{sec:fees}

\begin{figure*}[t]
\centering
\includegraphics[width=0.9\textwidth]{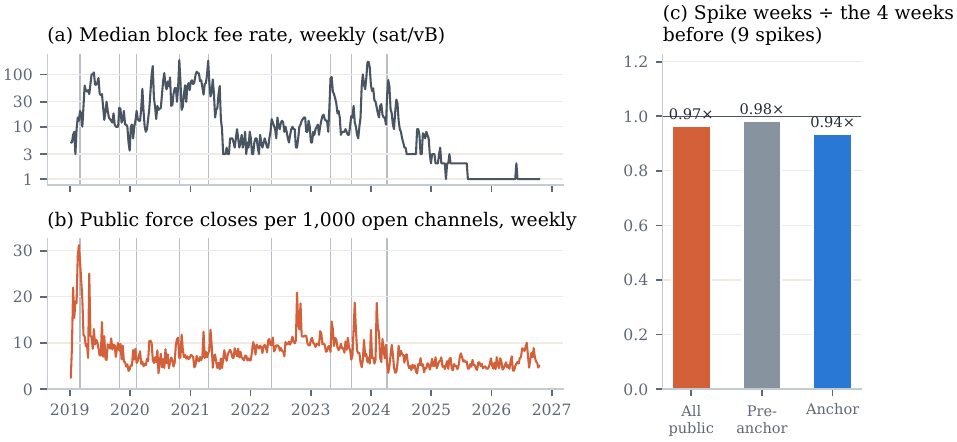}
\caption{Fee spikes and force closes. (a) Weekly median block fee rate (log scale); vertical lines mark the spike
weeks. (b) Weekly public force closes per 1,000 open channels, mass-close events excluded. (c) Force closes in the
spike week and the week after, divided by those in the four weeks before, pooled over the \NSpikes{} spikes.}
\label{fig:spikes}
\end{figure*}

\begin{table}[t]
\centering
\caption{Fee spikes against placebos. Top: organic public force closes in a window after the spike week, divided by
the four weeks before, pooled over the spikes, against random ordinary weeks of the same years. Bottom: elasticity of
the weekly organic force-close rate to the median block fee rate, within years, by the timing of the fee rate, against
placebos that move the fee series in time.}
\label{tab:spikes}
\footnotesize
\setlength{\tabcolsep}{3.5pt}
\begin{tabular}{lrrrr}
\toprule
Weeks after spike & Spikes & Ordinary & 90\% range & Pctl. \\
\midrule
All public, 0 to +1 & \PlEvAllNow & \PlEvAllNowMed & \PlEvAllNowLo--\PlEvAllNowHi & $\PlEvAllNowPct$ \\
Pre-anchor, 0 to +1 & \PlEvLegacyNow & \PlEvLegacyNowMed & \PlEvLegacyNowLo--\PlEvLegacyNowHi & $\PlEvLegacyNowPct$ \\
Anchor, 0 to +1 & \PlEvAnchorNow & \PlEvAnchorNowMed & \PlEvAnchorNowLo--\PlEvAnchorNowHi & $\PlEvAnchorNowPct$ \\
All public, +2 to +3\textsuperscript{a} & \PlEvAllLater & \PlEvAllLaterMed & \PlEvAllLaterLo--\PlEvAllLaterHi &
$\PlEvAllLaterPct$ \\
\midrule
Fee timing & $b$ & Null 95\% (C) & Pctl.\ (C) & Pctl.\ (B) \\
\midrule
1 week later (lead) & $\PlLeadOneB$ & $\PlLeadOneCircLo$ to $\PlLeadOneCircHi$ & $\PlLeadOneCircPct$ & $\PlLeadOneBlockPct$ \\
Same week & $\PlSameWeekB$ & $\PlSameWeekCircLo$ to $\PlSameWeekCircHi$ & $\PlSameWeekCircPct$ & $\PlSameWeekBlockPct$ \\
1 week earlier & $\PlLagOneB$ & $\PlLagOneCircLo$ to $\PlLagOneCircHi$ & $\PlLagOneCircPct$ & $\PlLagOneBlockPct$ \\
2 weeks earlier & $\PlLagTwoB$ & $\PlLagTwoCircLo$ to $\PlLagTwoCircHi$ & $\PlLagTwoCircPct$ & $\PlLagTwoBlockPct$ \\
3 weeks earlier & $\PlLagThreeB$ & $\PlLagThreeCircLo$ to $\PlLagThreeCircHi$ & $\PlLagThreeCircPct$ & $\PlLagThreeBlockPct$ \\
4 weeks earlier & $\PlLagFourB$ & $\PlLagFourCircLo$ to $\PlLagFourCircHi$ & $\PlLagFourCircPct$ & $\PlLagFourBlockPct$ \\
Largest, 1--4 earlier & $\PlMaxLagB$ & $\PlMaxLagCircLo$ to $\PlMaxLagCircHi$ & $\PlMaxLagCircPct$ & $\PlMaxLagBlockPct$ \\
\bottomrule
\end{tabular}
\par\smallskip\parbox{\linewidth}{\scriptsize Ordinary: median ratio over \PlWeekDraws{} draws of one ordinary week
per spike from the same year; 90\% range of those draws. Pctl.: percentile of the real value among the draws. (C):
circular shift of the fee series, all \PlShifts{} shifts; (B): \PlBlockWeeks-week blocks shuffled within each year,
\PlBlockDraws{} draws. \textsuperscript{a}Window chosen after seeing the lag profile.}
\end{table}

\subsection{Fee spikes do not bring a wave of force closes}

\paragraph{No wave}
Fee spikes did not bring a wave of force closes. Pooled over the \NSpikes{} spikes, organic public force closes in
the spike week and the week after were \PlEvAllNow{} times their level in the four weeks before
(Table~\ref{tab:spikes}). After random ordinary weeks of the same years the ratio is typically \PlEvAllNowMed, so the
spikes sit at percentile \PlEvAllNowPct{} of those draws, below a typical week rather than above it. A wave
\PlDetectableWavePct\% above the typical ratio would have been detected. Pre-anchor channels, whose fee is fixed at
signing and which an \texttt{update\_fee} disagreement would close in a spike, show the same (\PlEvLegacyNow), as do
anchor channels (\PlEvAnchorNow). Within years, the elasticity of the weekly force-close rate to the median fee rate
of the same week is $\PlSameWeekB$, at percentile \PlSameWeekCircPct{} of a placebo that moves the fee series in
time.

\paragraph{The spikes themselves}
The spikes went both ways. \NSpikesUp{} of the \NSpikes{} came with more force closes, the most in May~2023
(\SpikeMayTwentyThreePct\% more, mostly in anchor channels), when the median fee rate went to
\SpikeMayTwentyThreeFee~sat/vB from \SpikeMayTwentyThreeFeeBefore. Next came March~2019, with \SpikeMarNineteenPct\%
more, all in pre-anchor channels and in a month when our channel list may still have been filling in. The other
\NSpikesDown{} came with fewer force closes, as many as \SpikeAprTwentyFourPct\% fewer in April~2024
(Fig.~\ref{fig:spikes}).

\paragraph{No later rise either}
Closes broadcast during a spike at too low a fee would confirm only in the weeks after it, and the chain shows
confirmation, not broadcast. Those weeks show no rise beyond placebo noise. With the fee rate of one to
four weeks earlier, the elasticity is at most $\PlMaxLagB$; tested over the four lags together, $p=\PlFwPCirc$
against the circular shift and $p=\PlFwPBlock$ against the block shuffle. In the event windows, weeks +2 to +3 reach
\PlEvAllLater{} times the weeks before, at percentile \PlEvAllLaterPct{} of ordinary weeks.

\paragraph{Unlisted closes move the other way}
Unlisted force closes fell when fees rose (elasticity $\ElastUnlisted$). Mass closes cannot be removed from them,
because their nodes are unknown, so we do not interpret this.

\subsection{Fixed commitment fees overpaid in calm markets}

\begin{figure}[t]
\centering
\includegraphics[width=\columnwidth]{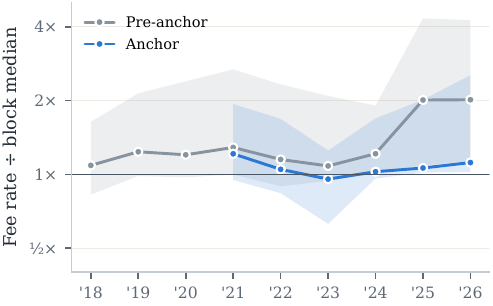}
\caption{A commitment's fee rate divided by the median fee rate of the block it confirmed in, by year: median and
interquartile range, pre-anchor and anchor commitments.}
\label{fig:guessing}
\end{figure}

\paragraph{Pre-anchor commitments paid twice the market when fees were low}
Pre-anchor commitments paid a median \LegacyMedRatioEarlyMin--\LegacyMedRatioEarlyMax{} times the median fee rate of
their own block from 2018 to 2024, and \LegacyMedRatioTwentySix{} times in 2025--26 (Fig.~\ref{fig:guessing}). In
2026, \LegacyOverTwoPctTwentySix\% paid more than double, against \LegacyOverTwoPctTwentyOne\% in 2021. Over the
period, pre-anchor commitments paid \LegacyOverBtcAll~BTC more than the tenth percentile of their blocks. A commitment
that paid too little waits for a cheaper block instead, so this measure shows overpayment; underpayment shows up as
delay.

\paragraph{Anchor commitments paid about the market}
Anchor commitments paid a median \AnchorMedRatioMin--\AnchorMedRatioMax{} times their block's median. In 2023,
\AnchorBelowPTenPctTwentyThree\% confirmed below their block's tenth percentile, that is, with a child's help.

\subsection{Lightning's closing bill is small}

{\setlength{\tabcolsep}{2.5pt}\begin{table}[t]
\centering
\caption{Lightning's closing bill: fees paid by channel closes and the transactions they require, public or not, against all Bitcoin transaction fees and block space.}
\label{tab:fee_share}
\footnotesize
\begin{tabular}{lrrrrr}
\toprule
Year & Coop.\ BTC & Force BTC & Total BTC & \% of fees & \% of block space \\
\midrule
2018\textsuperscript{b} & -- & 1.39 & -- & -- & -- \\
2019 & 1.74 & 5.44 & 7.18 & 0.04 & 0.04 \\
2020 & 2.12 & 5.14 & 7.25 & 0.03 & 0.03 \\
2021 & 2.19 & 4.91 & 7.10 & 0.03 & 0.05 \\
2022 & 0.82 & 3.29 & 4.11 & 0.08 & 0.12 \\
2023 & 2.40 & 14.43 & 16.83 & 0.07 & 0.10 \\
2024 & 1.88 & 9.55 & 11.43 & 0.08 & 0.08 \\
2025 & 0.28 & 2.09 & 2.37 & 0.14 & 0.10 \\
2026\textsuperscript{a} & 0.23 & 1.29 & 1.52 & 0.18 & 0.11 \\
\bottomrule
\end{tabular}
\par\smallskip\parbox{\linewidth}{\scriptsize \textsuperscript{a}January--October. Force: commitments, required spends (split per input) and third-party anchor sweeps; cooperative closes of channels outside our public list cannot be seen and are not included, nor are channel opens. \textsuperscript{b}From 12 January; cooperative closes of 2018 are mostly outside our list, so only force closes are shown.}
\end{table}
}

\paragraph{A fraction of a percent of Bitcoin's fees}
All closes and the spends they require, public or not, with third-party anchor sweeps, paid \BillBtcTwentyThree~BTC
in 2023, the most in any year, and \BillBtcTwentySix~BTC in the first ten months of 2026 (Table~\ref{tab:fee_share}). In
every year that was \BillFeePctMin--\BillFeePctMax\% of all transaction fees and at most \BillVbPctMax\% of block
space. The share of fees rose in the low-fee years because everything else became cheaper faster.

\section{Zero-Fee Commitments}
\label{sec:zerofee}

{\setlength{\tabcolsep}{3pt}\begin{table}[t]
\centering
\caption{Block space of the anchor mechanism per force close, observed (two P2WSH anchors and their spends) and under zero-fee commitments (one pay-to-anchor output spent by a child).}
\label{tab:zero_fee}
\footnotesize
\begin{tabular}{lrrrrr}
\toprule
Year & Anchor closes & Bumped \% & Anchors vB & P2A vB & Anchor BTC \\
\midrule
2021 & 3{,}839 & 8.4 & 194.1 & 155.9 & 0.022 \\
2022 & 34{,}297 & 12.2 & 200.4 & 151.8 & 0.207 \\
2023 & 39{,}280 & 36.6 & 180.5 & 124.6 & 0.231 \\
2024 & 41{,}035 & 12.6 & 163.1 & 151.3 & 0.226 \\
2025 & 59{,}927 & 2.9 & 166.0 & 162.0 & 0.339 \\
2026\textsuperscript{a} & 45{,}209 & 4.1 & 162.0 & 160.7 & 0.253 \\
\bottomrule
\end{tabular}
\par\smallskip\parbox{\linewidth}{\scriptsize \textsuperscript{a}Closes to 4 October 2026. Zero-fee closes that were not already bumped need a child of their own (wallet input and change output, 111.0 vB). Anchor BTC: value locked in 330-sat anchors, which pay-to-anchor outputs (0 sat) do not create.}
\end{table}
}

\paragraph{The first closes came before the specification}
The first zero-fee commitment we find closed a private channel on \ZfFirst, and \ZfBeforeMerge{} had closed before the
specification merged the format in May~2026. The first public one closed on \ZfFirstPublic. By the cutoff there were
\ZfCloses{}: \ZfUnlisted{} private and \ZfPublic{} public, involving \ZfPublicNodes{} distinct nodes, one of them party to
every public close. Every commitment paid no fee, and every pay-to-anchor output was spent by a child in the same
block, at a median package fee rate of \ZfPkgFr~sat/vB. The channels were small (median capacity \ZfMedCap~sat).

\paragraph{The new format saves little block space}
Across the \CfCloses{} anchor force closes we observe, the anchor mechanism took \CfAnchorVb~vB per close: two anchor
outputs and the transactions that spent them. Zero-fee commitments would have taken \CfPtwoaVb~vB: one pay-to-anchor
output, its spend, and a child for the \CfNotBumpedPct\% of closes that were not already bumped
(Table~\ref{tab:zero_fee}). That is \CfSavingPct\% less. The gains lie
elsewhere: the \CfAnchorBtc~BTC held in 330-sat anchor outputs would not exist, commitment fees no longer have to be
guessed at signing, and package relay replaces the carve-out that Bitcoin Core has
removed~\cite{core28,core31}.

\section{Discussion}
\label{sec:discussion}

\paragraph{Fee spikes do not break channels}
A common view holds that fee spikes break Lightning channels: commitment fees signed in calm markets fall behind,
peers disagree over \texttt{update\_fee}, and channels are forced closed in a wave. In confirmed force closes we see no
such wave, for pre-anchor channels as little as for anchor channels. \NSpikesUp{} of the \NSpikes{} spikes did bring
more force closes, the most in May~2023, but pooled they sit below ordinary weeks of the same years.
Spikes still matter, but through what happens after a close: HTLC transactions confirm later, owners bump commitments,
and sweepers wait. The caveat is that the chain records confirmation, not broadcast. A close broadcast in a spike at
too low a fee may wait for a cheaper block, which would spread a wave of broadcasts over the following weeks; those
weeks show no rise beyond placebo noise either.

\paragraph{Where the anchor design worked, and where it did not}
Anchor outputs were meant to let the broadcaster pay the fee the market asks at broadcast time. For commitments that
is rarely needed: outside 2023, owners seldom bumped, and in 2026 most same-block owner children did not raise the fee
rate at all. Where the design shows its value is in HTLC transactions. In fee spikes, anchor channels' HTLC timeouts
were late far less often than pre-anchor ones, and late HTLCs are what costs a forwarding node money. Meanwhile the
anchors themselves became third-party dust: owners leave them, sweepers collect them at the first block or wait for
cheap blocks, and miners take most of their value. This is the behaviour LND's open issue on dangling anchors
assumes~\cite{lnd8997}, measured, but only for P2WSH anchors. No third party sweeps simple-taproot anchors, and their
owners rarely do: of the \TaprootAnchorsTwentySix{} created in 2026, \TaprootAnchorsSpentTwentySix{} were spent. Each
taproot force close therefore leaves outputs in the UTXO set, which a zero-value pay-to-anchor output, spent together
with its commitment, does not (\S\ref{sec:zerofee}).

\paragraph{The migration}
Zero-fee commitments change little in block space. Their gains are a simpler fee story, no 330-sat outputs to sweep,
and relay rules that no longer depend on the removed carve-out. Their test is the one on which anchors improved on
the old format: whether HTLC transactions confirm before their deadlines when fees spike. Our measurements are the
baseline for it: at 50~sat/vB or more, \HtlcLateSpikeAnchor\% of anchor channels' HTLC timeouts confirmed more than
\HtlcMarginTight{} blocks late. How fast these gains arrive depends on operators,
not on the specification: a channel keeps its format until it closes. Outside one cluster of short-lived channels,
pre-anchor commitments were \LegacyPctExClusterTwentySix\% of 2026's public force closes, in channels a median of
\LegacyOldAgeYears~years old, years after anchors became available. Our second paper found that most Lightning
operators upgrade within months of a release while a tail stays far behind~\cite{kumar2026upgradedynamics}, so the old
formats are likely to stay on-chain for years. The same choice, between fixing a fee when a transaction is signed and paying it
when it is broadcast, faces every protocol built on pre-signed transactions and timelocks, and there too what matters
is whether the time-critical transactions confirm on time~\cite{riard2020timedilation}.

\paragraph{What a force close costs}
In fees, a force close is cheap and a channel's life costs a fraction of a percent of its capacity, and the open is a
large part of that. The cost of a force close is a ratio, a multiple of a cooperative close in fees and block space,
and a wait, days before the closer's own funds return; the wait outweighs the fees only in low-fee years. Its
exposure, the HTLCs that must confirm before their deadlines while fees are high, has rarely become a loss: since
\SplitResFromYear, forwarding nodes lost \SplitResForwarderPayments{} payments that way, each with the preimage already
on-chain.

\paragraph{What the chain can catch}
Two 2026 events show what transactions alone can reveal. A week of revoked commitments matched a splicing bug that
Core Lightning had fixed days before, and the only split resolutions in the order replacement cycling leaves were a
test of that attack on mainnet. We found both without node logs, and the same tests can be run on new blocks as they
arrive.

\section{Limitations}
\label{sec:limitations}
\emph{Broadcaster.} The chain does not show which party broadcast a force close, so costs are per channel, both
parties together.
\emph{Invisible bumps.} Fee bumps that never confirmed, and replaced transactions, leave no trace on the chain; we
have no mempool record.
\emph{Private channels.} Private channels appear only when they force close, so their number and their force-close
rate are unknown, and their cooperative closes are not in the closing bill. Before 2024, ``unlisted'' also includes
public channels missing from our list.
\emph{Open channels.} Channels whose nodes have left count as open until closed on-chain, which lowers the
per-channel rate as they accumulate; cohort incidence is not affected.
\emph{Penalties.} Only punished breaches are visible.
\emph{Scope of the cost account.} Pre-anchor \texttt{to\_remote} spends are excluded by design, and third-party
anchor sweeps are counted as value lost, not as fees the channel pays. The break-even rate counts the broadcaster's
locked balance only, not pending HTLC amounts. Simple taproot channels' spends are not
classified, so their delays, HTLCs and bumps are missing from \S\ref{sec:cost}; they were \UnlTaprootPctTwentySix\% of
2026's unlisted force closes.
\emph{Fee regimes.} Every result depends on the fee regime, which is why the window covers all of Lightning's history.
Anchor channels appear in our data from 2021, so anchor results start then.
\emph{Lifecycle.} Opens are costed under an attribution rule; dual-funded opens are not identified (which does not
change a per-channel cost), and only completed lives enter, so short-lived channels weigh more.
\emph{Client labels} come from our first paper, inherit its accuracy and start in May~2023.
\emph{Association.} The fee-spike test is a within-year association; it rules out a large same-week effect, not every
causal path.

\section{Ethics}
\label{sec:ethics}
All inputs are public blocks and public gossip. We report aggregates only: no list, ranking or label of nodes, sweeping
operators, addresses or transactions, in the paper or in the released data. In particular we do not identify the
dominant anchor sweeper, the nodes of the first zero-fee commitment closes, the operator of the replacement-cycling
test or the channels of the breach wave.
The paper reports no new vulnerability: replacement cycling was disclosed in 2023~\cite{riard2023replacement}, and the
splicing bug was fixed before the wave~\cite{cln26068}.

\section{Conclusion}
\label{sec:conclusion}
Fee spikes do not bring waves of Lightning force closes, for channels whose fee is fixed at signing or for those that
can bump it.
What a force close costs is a multiple of a cooperative close in fees and days of locked funds, and the wait outweighs
the fees only in low-fee years. When fees are high, pending payments are exposed, yet forwarders lost few of them, each
with the preimage already on-chain. The anchor design reduced that exposure for HTLCs, while its P2WSH anchors became
dust that third parties sweep and miners collect, and its taproot anchors stay in the UTXO set. Zero-fee commitments
will save little block space; the measure of the migration is whether payments still resolve before their deadlines
when the next spike comes.

\appendices
\section{Force Share by Client Pair}
\label{app:pairs}

\begin{table}[t]
\centering
\caption{Force share of public closes by the client pair at the two ends (closes from 8 May 2023).}
\label{tab:client_pairs}
\footnotesize
\begin{tabular}{lrrr}
\toprule
Client pair & Closes & Force & Force \% \\
\midrule
LND--LND & 169{,}435 & 47{,}905 & 28.3 \\
CLN--LND & 20{,}611 & 7{,}149 & 34.7 \\
LDK--LND & 17{,}749 & 6{,}603 & 37.2 \\
LDK--LDK & 13{,}415 & 204 & 1.5 \\
Eclair--LND & 11{,}474 & 3{,}335 & 29.1 \\
CLN--CLN & 1{,}832 & 740 & 40.4 \\
CLN--Eclair & 927 & 291 & 31.4 \\
\bottomrule
\end{tabular}
\par\smallskip\parbox{\linewidth}{\scriptsize Clients from paper 01's labels on the day of the close (high or medium confidence). The chain does not show which side broadcast a force close, so a close is attributed to the pair.}
\end{table}

Table~\ref{tab:client_pairs} gives the force share of public closes since 8~May~2023 by the implementations at the
two ends of the channel, from the labels of our first paper~\cite{kumar2026clientdiversity}. The chain does not show
which side broadcast a force close, so a close is attributed to the pair, not to one implementation. Among pairs with
both ends labelled and more than 1,000 closes, the force share ranges from \PairForcePctLndLnd\% (LND--LND,
\PairClosesLndLnd{} closes) to \PairForcePctClnCln\% (CLN--CLN); CLN--LND (\PairForcePctClnLnd\%), LDK--LND
(\PairForcePctLdkLnd\%) and Eclair--LND (\PairForcePctEclairLnd\%) lie between. Pairs with an unlabelled end run higher,
for example \PairForcePctLndUnknown\% for LND with an unlabelled peer. The one outlier, LDK--LDK at
\PairForcePctLdkLdkExact\% of \PairClosesLdkLdk{} closes, is one node closing channels cooperatively at scale, not a
property of the implementation: a single node is party to \PairLdkLdkTopNodePct\% of those closes. For the other
pairs we have not separated operators from implementations, so the table describes pairs and does not rank
implementations.

\section*{Data and Code Availability}
Code, the aggregate datasets behind every figure and table, and a script that recomputes every number quoted in the
text are available at \url{https://github.com/bitcoindatalabs/ln-force-close-costs}. All stages truncate their
inputs at the data cutoff, block \CutoffHeight{} (\CutoffDate). The released aggregates name no node, address or
transaction. The underlying blocks are public; our channel list and the per-transaction tables are available from
the author on request.

\bibliographystyle{IEEEtran}
\bibliography{references}

\begin{thebibliography}{10}
\providecommand{\url}[1]{#1}
\csname url@samestyle\endcsname
\providecommand{\newblock}{\relax}
\providecommand{\bibinfo}[2]{#2}
\providecommand{\BIBentrySTDinterwordspacing}{\spaceskip=0pt\relax}
\providecommand{\BIBentryALTinterwordstretchfactor}{4}
\providecommand{\BIBentryALTinterwordspacing}{\spaceskip=\fontdimen2\font plus
\BIBentryALTinterwordstretchfactor\fontdimen3\font minus
  \fontdimen4\font\relax}
\providecommand{\BIBforeignlanguage}[2]{{%
\expandafter\ifx\csname l@#1\endcsname\relax
\typeout{** WARNING: IEEEtran.bst: No hyphenation pattern has been}%
\typeout{** loaded for the language `#1'. Using the pattern for}%
\typeout{** the default language instead.}%
\else
\language=\csname l@#1\endcsname
\fi
#2}}
\providecommand{\BIBdecl}{\relax}
\BIBdecl

\bibitem{poon2016bitcoin}
J.~Poon and T.~Dryja, ``The {B}itcoin {L}ightning {N}etwork: Scalable off-chain
  instant payments,''
  \url{https://lightning.network/lightning-network-paper.pdf}, 2016, draft
  version 0.5.9.2, January 14, 2016.

\bibitem{bolt3}
{Lightning Network Developers}, ``{BOLT \#3: Bitcoin Transaction and Script
  Formats},''
  \url{https://github.com/lightning/bolts/blob/master/03-transactions.md},
  accessed 2026-10-07.

\bibitem{bolt5}
------, ``{BOLT \#5: Recommendations for On-chain Transaction Handling},''
  \url{https://github.com/lightning/bolts/blob/master/05-onchain.md}, accessed
  2026-11-10.

\bibitem{harris2020flood}
J.~Harris and A.~Zohar, ``Flood \& loot: A systemic attack on the {L}ightning
  {N}etwork,'' in \emph{Proceedings of the 2nd ACM Conference on Advances in
  Financial Technologies (AFT '20)}.\hskip 1em plus 0.5em minus 0.4em\relax
  ACM, 2020, pp. 202--213.

\bibitem{riard2020timedilation}
A.~Riard and G.~Naumenko, ``Time-dilation attacks on the {L}ightning
  {N}etwork,'' arXiv:2006.01418, 2020.

\bibitem{riard2023replacement}
A.~Riard, ``Full disclosure: {CVE-2023-40231} / {CVE-2023-40232} /
  {CVE-2023-40233} / {CVE-2023-40234} ``all your mempool are belong to us'',''
  bitcoin-dev mailing list,
  \url{https://mirror.b10c.me/lists/bitcoindev/CALZpt+GdyfDotdhrrVkjTALg5DbxJyiS8ruO2S7Ggmi9Ra5B9g@mail.gmail.com},
  Oct. 2023, posted 2023-10-16. Accessed 2026-10-15.

\bibitem{core31}
{Bitcoin Core}, ``Bitcoin core 31.0 release notes,''
  \url{https://bitcoincore.org/en/releases/31.0/}, Apr. 2026, accessed
  2026-11-10.

\bibitem{bolts1228zerofee}
B.~Teinturier, ``Zero-fee commitments using v3 transactions (feature 40/41),''
  lightning/bolts pull request \#1228,
  \url{https://github.com/lightning/bolts/pull/1228}, 2026, opened 2025-02-14,
  merged 2026-05-04. Accessed 2026-11-10.

\bibitem{bip431}
G.~Zhao, ``{BIP} 431: Topology restrictions for pinning,''
  \url{https://github.com/bitcoin/bips/blob/master/bip-0431.mediawiki}, 2024,
  accessed 2026-11-10.

\bibitem{core28}
{Bitcoin Core}, ``Bitcoin core 28.0 release notes,''
  \url{https://bitcoincore.org/en/releases/28.0/}, Oct. 2024, accessed
  2026-11-10.

\bibitem{bitmex2020closures}
{BitMEX Research}, ``Lightning network (part 6) -- over 60,000 non-cooperative
  channel closures,'' BitMEX Blog,
  \url{https://blog.bitmex.com/lightning-network-part-6-over-60000-non-cooperative-channel-closures/},
  Jan. 2020, posted 2020-01-13. Archived copy accessed 2026-11-10.

\bibitem{zabka2022empirical}
P.~Zabka, K.-T. F{\"o}rster, S.~Schmid, and C.~Decker, ``Empirical evaluation
  of nodes and channels of the {L}ightning {N}etwork,'' \emph{Pervasive and
  Mobile Computing}, vol.~83, p. 101584, 2022.

\bibitem{grotschla2025lifecycle}
F.~Gr{\"o}tschla, L.~Heimbach, S.~Richner, and R.~Wattenhofer, ``On the
  lifecycle of a {L}ightning {N}etwork payment channel,'' in \emph{Financial
  Cryptography and Data Security. FC 2025 International Workshops, Part I},
  ser. Lecture Notes in Computer Science.\hskip 1em plus 0.5em minus
  0.4em\relax Springer, 2025, pp. 1--16.

\bibitem{antonelli2026closures}
S.~Antonelli, V.~Davis, H.~Rush, A.~Potdevin, J.~Shrader, V.~Singh, and
  E.~Rossi, ``Predicting channel closures in the {Lightning Network} with
  machine learning,'' arXiv:2605.12759, 2026, accepted at IEEE BCCA 2026.

\bibitem{daftuar2024v3}
S.~Daftuar, ``Analysis of attempting to imbue {LN} commitment transaction
  spends with v3 semantics,'' Delving Bitcoin,
  \url{https://delvingbitcoin.org/t/analysis-of-attempting-to-imbue-ln-commitment-transaction-spends-with-v3-semantics/527},
  Feb. 2024, posted 2024-02-07. Accessed 2026-11-10.

\bibitem{lnd8997}
M.~Morehouse, ``Remove dangling anchor sweeping,'' lightningnetwork/lnd issue
  \#8997, \url{https://github.com/lightningnetwork/lnd/issues/8997}, Aug. 2024,
  opened 2024-08-09. Accessed 2026-11-10.

\bibitem{kumar2026clientdiversity}
\BIBentryALTinterwordspacing
S.~Kumar, ``Does client diversity protect the {L}ightning {N}etwork? {E}vidence
  from the 2026 {C}ore {L}ightning shutdown,'' arXiv:2610.20952 [cs.CR], Oct.
  2026. [Online]. Available: \url{https://arxiv.org/abs/2610.20952}
\BIBentrySTDinterwordspacing

\bibitem{kumar2026upgradedynamics}
\BIBentryALTinterwordspacing
------, ``Who patches the {L}ightning {N}etwork? {E}vidence from release lines
  in gossip, 2023--2026,'' arXiv:2610.30350 [cs.CR], Oct. 2026. [Online].
  Available: \url{https://arxiv.org/abs/2610.30350}
\BIBentrySTDinterwordspacing

\bibitem{bolt7}
{Lightning Network Developers}, ``{BOLT \#7: P2P Node and Channel Discovery},''
  \url{https://github.com/lightning/bolts/blob/master/07-routing-gossip.md},
  accessed 2026-10-07.

\bibitem{bolt2}
------, ``{BOLT \#2: Peer Protocol for Channel Management},''
  \url{https://github.com/lightning/bolts/blob/master/02-peer-protocol.md},
  accessed 2026-10-07.

\bibitem{bolts688anchors}
J.~Jager, ``Anchor outputs,'' lightning/bolts pull request \#688,
  \url{https://github.com/lightning/bolts/pull/688}, 2020, opened 2019-10-28,
  merged 2020-08-19. Accessed 2026-11-10.

\bibitem{bolts995taproot}
O.~Osuntokun, ``extension-bolt: simple taproot channels (feature 80/81),''
  lightning/bolts pull request \#995,
  \url{https://github.com/lightning/bolts/pull/995}, 2026, opened 2022-05-30,
  merged 2026-05-04. Accessed 2026-11-10.

\bibitem{optechpinning}
{Bitcoin Optech}, ``Transaction pinning,'' Bitcoin Optech topics,
  \url{https://bitcoinops.org/en/topics/transaction-pinning/}, accessed
  2026-11-10.

\bibitem{kappos2021empirical}
G.~Kappos, H.~Yousaf, A.~Piotrowska, S.~Kanjalkar, S.~Delgado-Segura,
  A.~Miller, and S.~Meiklejohn, ``An empirical analysis of privacy in the
  {L}ightning {N}etwork,'' in \emph{Financial Cryptography and Data Security
  (FC 2021), Part I}, ser. Lecture Notes in Computer Science, vol. 12674.\hskip
  1em plus 0.5em minus 0.4em\relax Springer, 2021, pp. 167--186.

\bibitem{romiti2021crosslayer}
M.~Romiti, F.~Victor, P.~Moreno-Sanchez, P.~S. Nordholt, B.~Haslhofer, and
  M.~Maffei, ``Cross-layer deanonymization methods in the {L}ightning
  protocol,'' in \emph{Financial Cryptography and Data Security (FC 2021), Part
  I}, ser. Lecture Notes in Computer Science, vol. 12674.\hskip 1em plus 0.5em
  minus 0.4em\relax Springer, 2021, pp. 187--204.

\bibitem{sguanci2023massexit}
C.~Sguanci and A.~Sidiropoulos, ``Mass exit attacks on the {L}ightning
  {N}etwork,'' in \emph{2023 IEEE International Conference on Blockchain and
  Cryptocurrency (ICBC)}.\hskip 1em plus 0.5em minus 0.4em\relax IEEE, 2023,
  pp. 1--3.

\bibitem{mizrahi2021congestion}
A.~Mizrahi and A.~Zohar, ``Congestion attacks in payment channel networks,'' in
  \emph{Financial Cryptography and Data Security (FC 2021), Part II}, ser.
  Lecture Notes in Computer Science, vol. 12675.\hskip 1em plus 0.5em minus
  0.4em\relax Springer, 2021, pp. 170--188.

\bibitem{harding2024irrevocable}
D.~A. Harding, ``Disclosure: irrevocable fees---stealing from {LN} using
  revoked commitment transactions,'' Delving Bitcoin,
  \url{https://delvingbitcoin.org/t/disclosure-irrevocable-fees-stealing-from-ln-using-revoked-commitment-transactions/1314},
  Dec. 2024, posted 2024-12-11. Accessed 2026-10-15.

\bibitem{mempoolspace}
{mempool.space}, ``Lightning network explorer: channel close types,''
  \url{https://mempool.space/lightning}, accessed 2026-10.

\bibitem{lnresearchtopology}
{lnresearch}, ``topology: Data about the past and current structure of the
  {L}ightning {N}etwork,'' \url{https://github.com/lnresearch/topology},
  accessed 2026-11-04.

\bibitem{cln26068}
{Core Lightning}, ``Core lightning v26.06.8 release: ``a unilateral close after
  a splice locked could broadcast a revoked commitment, losing the channel
  funds to a penalty'',''
  \url{https://github.com/ElementsProject/lightning/releases/tag/v26.06.8},
  Sep. 2026, released 2026-09-19; fix in commit c65eec307 (2026-09-14).
  Accessed 2026-11-11.

\bibitem{boltsimpletaproot}
{Lightning Network Developers}, ``Extension {BOLT}: Simple taproot channels,''
  \url{https://github.com/lightning/bolts/blob/191462825ae341f0a27098765c075f01ca9bc6c2/bolt-simple-taproot.md},
  2026, bolt-simple-taproot.md at commit 1914628 (2026-10-19), added by pull
  request \#995. Accessed 2026-11-10.

\end{thebibliography}

\end{document}